\documentclass[twocolumn]{aastex631}

\accepted{\today}

\submitjournal{APJ}

\usepackage{graphicx}
\usepackage[caption=false]{subfig}

\begin{document}

\title{
  Investigating Extended Main-Sequence Turnoffs in Galactic Open Clusters
       }

\correspondingauthor{Khushboo K. Rao}
\email{khushboo@astro.ncu.edu.tw}

\author[0000-0001-7470-9192]{Khushboo K. Rao}
\affiliation{Institute of Astronomy, National Central University, 320317 Taoyuan, Taiwan}

\author[0000-0003-0262-272X]{Wen Ping Chen}
\affiliation{Institute of Astronomy, National Central University, 320317 Taoyuan, Taiwan}
\affiliation{Department of Physics,  National Central University, 320317 Taoyuan, Taiwan}


\begin{abstract}

The extended main sequence (eMS) and extended main sequence turnoff (eMSTO) phenomena have been observed in some young and intermediate-age star clusters in the Milky Way and in the Magellanic Clouds. In this study, we conduct a survey of 53 galactic open clusters (OCs) to investigate the roles of stellar rotation, differential extinction, and cluster properties in the emergence of eMS and eMSTO. The projected rotational velocities are taken from the \textit{Gaia} ESO spectroscopic survey and the \textit{Gaia} DR3 line-broadening velocities. Stellar members of each OC are identified using the ML-MOC algorithm with \textit{Gaia} DR3 astrometry. 
We divide clusters into four classes based on the color-rotation distribution, extinction, and MSTO morphology and report 14 clusters (Class~I) that exhibit split MS with fast and slow rotators populating the redder and bluer parts of MSTO. For the remaining clusters, differential extinction hampers the color-rotation distinction and also inflates MSTO width and therefore introduces a systematic offset in the MSTO-age relation. We also quantify the fraction of slow rotators among MSTO stars, finding a median value of $f_{\rm slow\, rot}^{v \sin i<100} \approx 0.41$ and the fraction reaching the spin-down limit, $f_{\rm slow\, rot}^{v \sin i<30}$, is  $ \approx 0.08$. We find no statistically significant correlation between $f_{\rm slow\, rot}$ and either the binary fraction or cluster age.
\end{abstract}


\section{Introduction} \label{sec:intro}

Extended main sequence turnoffs \citep[eMSTOs;][]{Mackey2007} and extended main sequences \citep[eMSs;][]{Cordoni2018} are observed in several young and intermediate-age clusters (ages $\leq 2.0$~Gyr), for which their upper MS and main sequence turnoff (MSTO) regions appear significantly broader than the rest of their MS, subgiant branches, and red giant branches. The phenomena have attracted special attention in the last two decades since the detection of double MSTO stars in the Large Magellanic Cloud (LMC) cluster, NGC\,1846 \citep{Mackey2007}. Subsequently, the eMSTO/eMS phenomenon was observed in several MC clusters younger than 1 to 2~Gyr old \citep{Mackey2008, Glatt2008, Milone2009, Milone2015, Goudfrooij2011, Li2014, Correnti2017}. A study based on \textit{Gaia} Data Release 2 \citep[DR2;][]{GaiaDR22018} revealed that eMSs/eMSTOs are not unique to LMC and SMC clusters but are also common among open clusters (OCs) of young and intermediate ages in the Milky Way galaxy \citep{Cordoni2018}. 

Different stellar rotation rates of MSTO stars are currently accepted as the primary channel for these features \citep{Bastian2009, Bastian2018, Marino2018, Marino2018b}. The upper MS and MSTO regions in CMDs of these clusters are populated by intermediate-mass stars ($\sim$1.3--8 M$_\odot$, corresponding to B to early F spectral types), which have radiative envelopes. Therefore, their fast stellar rotation is supported by the fact that stars spin up as they contract toward the zero-age main sequence due to the conservation of angular momentum and retain most of their initial angular momentum throughout the MS phase, unless they join the zero-age MS rotating near the critical limit \citep{Gagnier2019A&A...625A..89G}. The rapid rotation produces a gravity-darkening effect, which affects the observed effective temperature of the star based on its inclination \citep{Espinosa2011A&A...533A..43E}. In addition, part of the stellar energy budget is channeled into rotation, thus reducing the luminosity of fast rotators relative to non-rotating stars \citep{Maeder2000, Ekstrom2012A&A...537A.146E}. 

However, several spectroscopic studies and simulations of individual clusters demonstrated that MSTO stars of intermediate- and young-age clusters possess slow- to moderate-rotators as well \citep{Li2019, Kamann2020, Cordoni2024, Deng2024}. This posed a challenge as to what causes the varying degree of angular momentum loss in stars with radiative envelopes. Binary evolution \citep{Kamann2021}, stellar disk interactions in the pre-MS phase \citep{Bastian2020}, and stellar mergers \citep{Wang2022} are currently being explored to explain the presence of the slow rotators. The majority of slow rotators typically occupy the bluer regions of CMDs \citep{Cordoni2024}. Thus, together these slow and fast rotators naturally generate the broad distributions in temperature and luminosity characteristic of eMSs/eMSTOs.

In addition to differential rotation, a varying degree of line-of-sight extinction across a cluster can also contribute to MSTO width and displace fast and slow rotators. To address this issue, a common approach is to perform differential reddening correction \citep{Platais2012ApJ...751L...8P, Cordoni2018, Souza2025A&A...701A.221S}. Methods based on cluster CMD and spatial distribution work well for high-density clusters but are less effective for low-density clusters due to the loss of spatial resolution. Additionally, available dust maps often are limited in angular resolution, making cluster-scale differential extinction difficult to assess. The position of cluster members on the CMD, combined with their rotational velocities, is closely linked to their evolutionary histories \citep{Bastian2025A&A...700A.241B, Mathieu2025ARA&A..63..467M}. Therefore, clusters experiencing significant line-of-sight extinction may have stars misplaced upon reddening correction, potentially altering the interpretation of stellar populations, which cannot be reliably resolved without spectroscopic constraints. Furthermore, as stellar rotation models continue to be developed \citep{Nguyen2022, Nguyen2025A&A...701A.258N}, they should be tested against clusters minimally affected by varying extinction to ensure robust calibration. Such testing will help determine whether very fast rotators located in the redder region of MSTO can be explained by rotating stellar evolution models alone or require other explanations like past/ongoing interactions.

Star clusters are dynamic environments where stellar populations are subjected to stellar interactions, and thus the combined effect of cluster density and binary/multiplicity fraction can alter the stellar population and their properties \citep{Boffin2015ASSL..413.....B, Mathieu2025ARA&A..63..467M}.  For example, among fast rotators, several Be, B[e], and shell stars have been observed that possess decretion disks \citep{Kamann2023MNRAS.518.1505K}. Currently, many of these type of stars are known to form through past or ongoing stellar interactions in binary systems \citep{Li2026ApJ...996L..42L,Rivinius2026enap....2..430R}. Additionally, intermediate-mass stars form alongside massive O-type stars, which have high-energy stellar winds that can evaporate or truncate the circumstellar disks surrounding these intermediate-mass pre-MS stars. This can reduce the likelihood of their star-disk interactions and increase the fraction of fast rotators. Therefore, the interplay of cluster density and binary/multiple fraction, along with cluster age, can result in varying fractions of slow and fast rotators, even within coeval star clusters. To effectively calibrate future stellar rotation models and understand the formation of fast and slow rotators, it is essential to categorize clusters based on their properties, such as degree of line-of-sight extinction, CMD features, and number of MSTO stars.

In this study, we examined MSTO morphologies of 53 OCs and classified them into different categories based on their extinction and CMD qualities to single out clusters suitable for reliable physical interpretation. Following this classification, we also quantified the possible correlation between the occurrence of eMS/eMSTO and projected rotational velocities ($v\sin i$) distribution and various cluster parameters such as age, mass, extinction, or metallicity. We further estimate the fraction of slow rotators among MSTO stars, examine its dependence on cluster age and binary fraction, and assess the specific contributions of known binaries, pulsating variables, and chemically peculiar stars to the observed slow-rotator population.

The remainder of the paper is organized as follows. In \S\ref{sec:data}, we describe the data, cluster membership determination, differential reddening correction, isochrone fitting, and binary fraction estimation. In \S\ref{sec:analysis}, we present the cluster classification, analyze the contributing factors to MSTO width, and investigate the fraction of slow rotators and the role of binaries and variables. In \S\ref{sec:summary}, we summarize our findings and conclude the results.
 
\section{Data and Analysis }\label{sec:data}

\begin{figure*}
    \centering
\includegraphics[width = 0.35\textwidth]{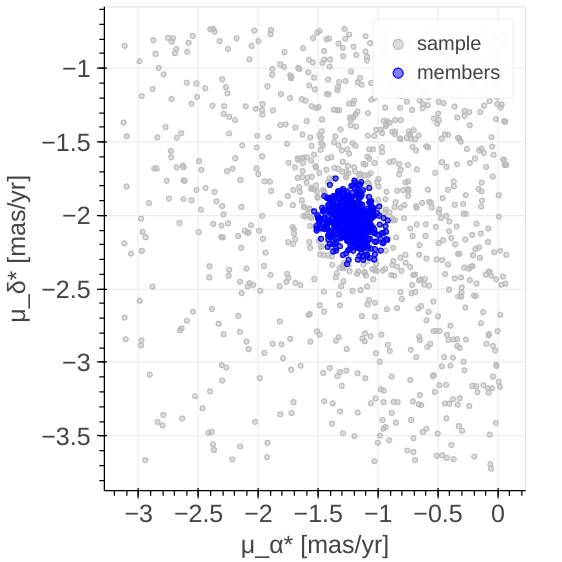}
 \includegraphics[width = 0.35\textwidth]{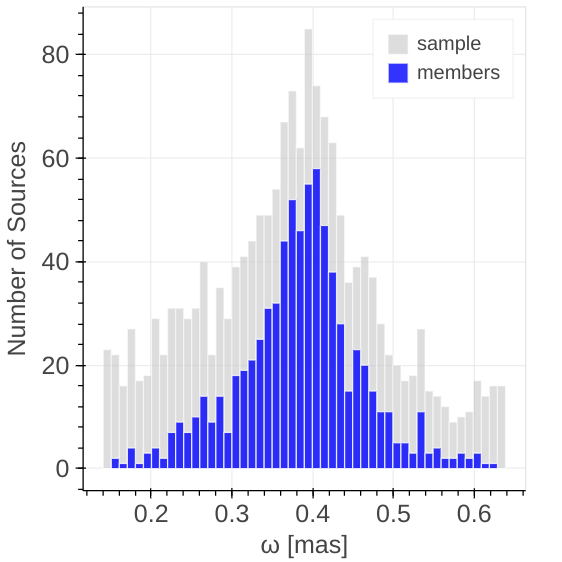}
\includegraphics[width = 0.35\textwidth]{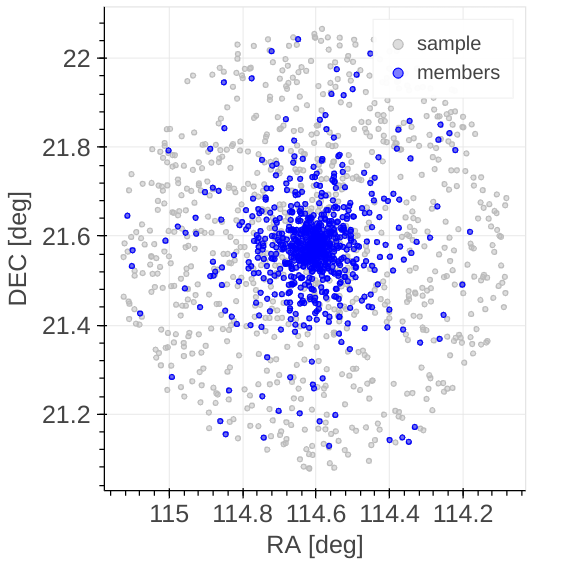} 
\includegraphics[width = 0.35\textwidth]{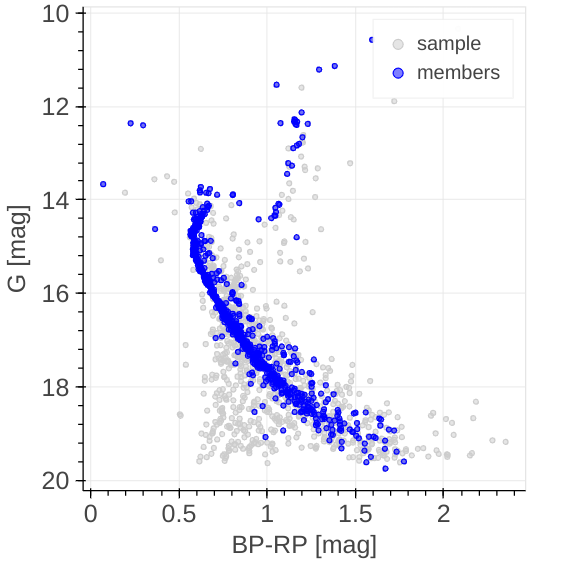}
    \caption{NGC\,2420 as an example of the selection of member stars (blue) and sample sources (grey) using the ML-MOC algorithm.}
    \label{fig:membership}
\end{figure*}

For this work, we selected OCs based on the availability of either of the $v\sin i$ from the Gaia-ESO Public Spectroscopic Survey \citep[GES;][]{Randich2022} and the line-broadening velocities ($v_{\text{broad}}$) from \textit{Gaia} DR3 \citep{GaiaDR32023}.

GES is a ground-based spectroscopic survey targeting $\sim10^5$ stars in the Milky Way galaxy, with an emphasis on OCs. During a 15-year timeline, spectroscopic data have been collected using both VLT/GIRAFFE and VLT/UVES \citep{Gilmore2022,Randich2022}. The GES survey covers OCs of a wide range of evolution, from $\sim 1$~Myr to $\sim 8$~Gyr, sampling different environments and star formation conditions.  The data provide stellar parameters, abundances, and velocities for more than 110000 stars. In this work, we use $v \sin i$ measurements of the members of our target OCs when available.   

To select OCs with GES $v\sin i$ measurements, we refer to Table A.1 of \citet{Bragaglia2022}, which lists young and intermediate-age massive clusters observed by the survey. Initially, 29 OCs with ages between 10~Myr and 2~Gyr were selected. Of these, 15 were excluded due to an insufficient number of MSTO stars with $v \sin i$ measurements. An additional cluster, NGC\,6281, was removed due to its sparse population and significant field star contamination, preventing a robust membership determination. This produced a final sample of 13 OCs with $v \sin i$ measurements from the GES data.

To supplement the GES sample, we used the \textit{Gaia} DR3 $v_{\text{broad}}$ measurements for stellar rotation, adopting the OC catalog by \citet{Cavallo2024}.  The \textit{Gaia} DR3, in addition to providing astrometric and photometric data for billions of sources, also provides $v_{\text{broad}}$ for more than 3.5 million stars that have $T_{\text{eff}}$ between 3500~K and 14500~K and G$_{\text{RVS}}$ brighter than $\approx12$~mag \citep{Fremat2023}. The $v_{\text{broad}}$ values are derived using Ca~II triplets from spectra obtained from the Radial Velocity Spectrometer ($\Delta \lambda/\lambda\sim11,500$) onboard the \textit{Gaia} spacecraft, which operates in the 846~nm to 870~nm wavelength range. Several factors contribute to spectral line broadening in addition to stellar rotation, e.g., macroscopic random motions, prominences, radial and nonradial pulsations, systematic velocity fields related to stellar winds, undetected binarity, limited accuracy of the line spread function, and stray light correction. \citet{Fremat2023} reported that $v_{\text{broad}}$ of \textit{Gaia} DR3 serves as a proxy for $v \sin i$ only within specific $G_{\text{RVS}}$ and $T_{\text{eff}}$ domains.  However, by comparing \textit{Gaia} DR3 $v_{\text{broad}}$  with $v \sin i$ from the GALAH DR3 \citep{Buder2021} and \citet{Glcebocki2005}, \citet{Cordoni2024} demonstrated that $v_{\text{broad}}$ derived from \textit{Gaia} DR3 can be reliably used as proxies for $v \sin i$. 

Because the $v_{\text{broad}}$ measurements from the \textit{Gaia} DR3 data are available only for bright ($G \lesssim 13$~mag) stars, we selected 40 OCs that each have a sufficient number of MSTO and upper MS stars with $v_{\text{broad}}$ measurements.

Combining both data sets led to a final sample of 53 OCs: 13 with $v \sin i$ from GES and 40 with $v_{\text{broad}}$ from \textit{Gaia} DR3. These OCs cover a broad age range, from $\sim10$~Myr to $\sim2.4$~Gyr. Younger clusters in the sample are particularly useful for investigating the possible early emergence of the eMS and eMSTO features, whereas older ones are used to explore the possible disappearance of these signatures and to understand the physical mechanisms responsible for the evolutionary transition. 

\subsection{Cluster Membership}\label{membership}

To identify members of the OCs, we used the ML-MOC algorithm \citep{Agarwal2021} on \textit{Gaia} DR3 data. The ML-MOC algorithm is a machine learning-based approach to determine membership in an OC. It employs two unsupervised algorithms: the Gaussian Mixture Model (GMM) and k-Nearest Neighbors (kNN). This algorithm utilizes parallax and proper motion measurements from the \textit{Gaia} data without relying on prior information about the cluster. Gauged by radial velocity measurements, membership identified by the ML-MOC algorithm has a contamination fraction of 2\% to 12\% \citep{Agarwal2021,Bhattacharya2022}, and member identification is robust, being 90\% complete down to $G = 19$~mag \citep{Bhattacharya2022}. The ML-MOC algorithm has been productive in the discovery of, for example, tidal tails \citep{Bhattacharya2021,Bhattacharya2022}, exotic stellar populations \citep{Rao2022,Rao2023Aplus}, and double blue straggler star sequences \citep{Rao2023}.  Of the sample of 53 OCs (\S\ref{sec:data}), 22 OCs are in common with \citet{Bhattacharya2022} and 2 OCs with \citet{Rao2023}. We adopted the members of these 24 OCs.  For the remaining 28 OCs, we used the ML-MOC algorithm to identify members. We describe below the methodology of member identification using NGC\,2420 as an example, illustrated in Figure~\ref{fig:membership}.

\begin{figure*}
    \centering
    \includegraphics[width=\textwidth]{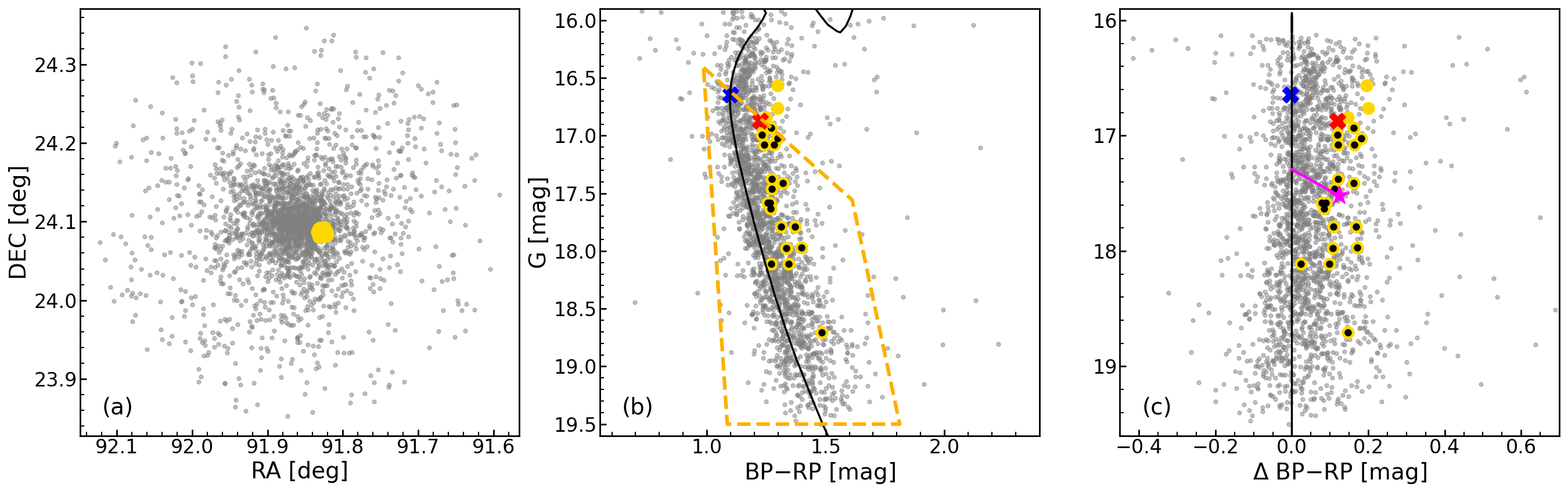}
    \includegraphics[width=\textwidth]{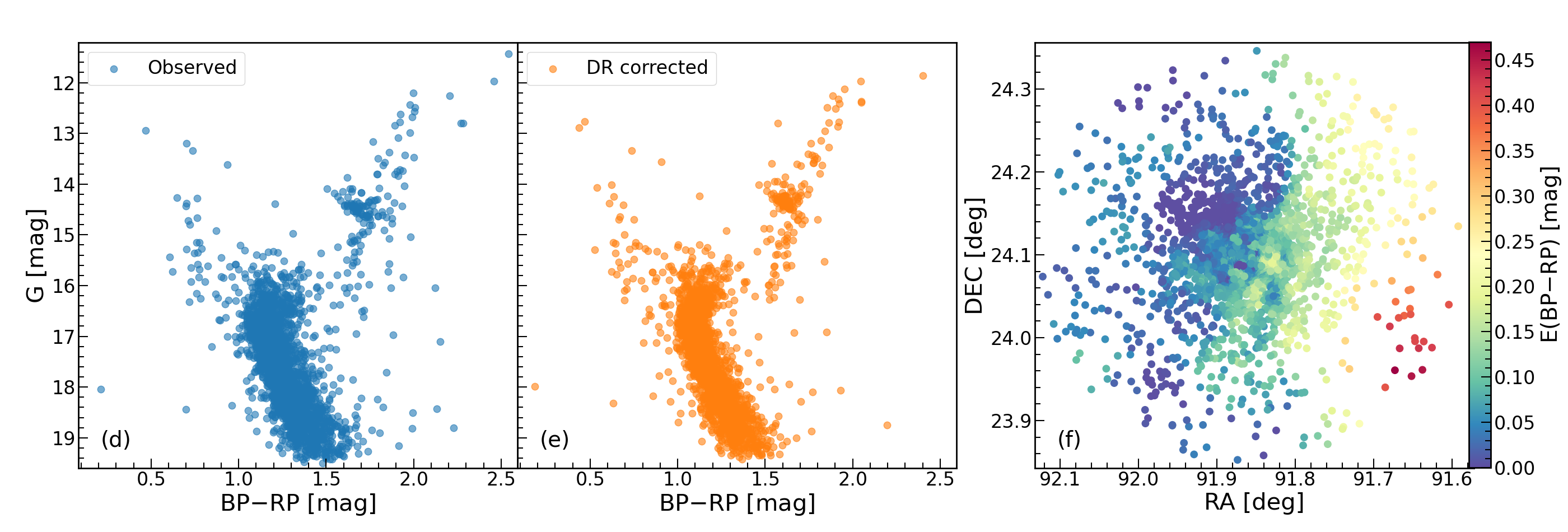}
    \caption{The correction of differential reddening illustrated by NGC\,2158. (a)~Spatial distribution of the cluster members (in grey) and of the 25 nearest neighbors (in orange) of a sample star. (b)~The CMD with PARSEC isochrone (black solid line) as a reference line, selection box (orange dashed lines), and nearest neighbors (orange dots) for the differential reddening correction. Black dots show members within the box, which are used to compute differential reddening and extinction. The star being corrected is depicted as a red cross, and its corrected position as a blue cross. (c)~Same as (b), but with isochrone-normalized colors. (d)~The observed \textit{Gaia} CMD. (e)~The CMD as in (d), but differential reddening corrected. (f)~The reddening map across the cluster.}
    \label{fig:DR_ngc2158_demo}
\end{figure*}

First, we selected all \textit{Gaia} DR3 sources within a 30 arcmin field of the cluster that have positions, proper motions, parallaxes, $G$~mag, $G_{\text{BP}}$~mag, $G_{\text{RP}}$~mag, and non-negative parallaxes with errors of $G < 0.05$~mag and are called \textit{All}~\textit{Sources}. kNN is then applied to parallaxes and proper motions of \textit{All}~\textit{Sources} to remove obvious field stars and create a sample of sources that have fewer field stars compared to the cluster members. We call these sources \textit{Sample}~\textit{Sources}. GMM is then applied to parallaxes and proper motions of \textit{Sample} \textit{Sources} to find possible cluster members. GMM disentangles cluster members from the field stars and assigns membership probability to each source. Here, we only selected the cluster members with membership probability $> 0.6$ to minimize the field star contamination. We included the cluster members with membership probability from 0.2 to 0.6 by increasing the range of the proper motion while keeping the parallax fixed as the parallax of sources with membership probability $> 0.8$. The proper motion, parallax, spatial distributions, and CMD of sample sources and the identified cluster members for NGC\,2420 are shown in Fig.~\ref{fig:membership}. A total of 560 members are identified within 30 arcmin of the cluster center. We estimated the cluster center using the mean-shift algorithm \citep{Comaniciu_meanshift} and determined the cluster radius as 15 arcmin, beyond which the cluster members were indistinguishable from the field stars. Finally, we have 512 members in total within the 15-arcmin cluster radius. The OCs used in this study, their central coordinates, and radii are listed in Table~\ref{tab:fundamental_params}. 


\subsection{Differential reddening correction} \label{DR_corr}

Being located near the galactic plane, OCs are often affected by varying levels of interstellar extinction, much more so for young systems physically associated with parental gas and dust. 
This results in a broadening of their CMDs, leading to an inaccurate estimation of the fundamental parameters, such as age, metallicity, and distance, determined through isochrone fitting.  The extinction, sometimes patchy, can create a false appearance of a double MS, an erroneous identification of highly reddened stars as binary stars, etc. 

To address this issue, we performed differential reddening corrections to OCs following the method presented in \citet{Rao2023Aplus}, here demonstrated with NGC\,2158 as an example.  We first fitted a PARSEC isochrone to the cluster's CMD by adopting the fundamental parameters in \citet{Rao2023Aplus}. The fitted isochrone served as a fiducial reference for the differential reddening correction. We adopted a reddening vector $R_{\text{G}} = 1.875$  \citep{Rao2023Aplus}, which was well-fitted along the distortion of the red clumps of intermediate-age OCs. We used only MS stars to correct for differential reddening. For this, we created an MS grid using $R_{\text{G}}$ as the upper limit, the magnitude of the faintest star in the $G$ band as the lower limit, and setting lateral boundaries to enclose the MS.

We then selected the 25 nearest neighbors of each star in the spatial distribution using the kNN algorithm. The number 25 is empirical to maintain a sufficient number of members on the outskirts of a cluster, where the member density decreases. Of these 25 stars, we selected the ones located within the MS grid to find the extinction/reddening of the star. We then normalized the $BP-RP$~mag of stars within the MS grid using the fitted isochrone and estimated the average of their normalized $BP-RP$ and $G$~mags. 

The differential reddening and extinction of the star are obtained by tracing this average data point to the normalized isochrone along the reddening vector $R_{\text{G}}$, and are used thereby to correct $G$ and $BP-RP$ of the star. These steps, depicted in Fig.~\ref{fig:DR_ngc2158_demo} for NGC\,2158, are applied to correct $G$ and $BP-RP$ of each cluster member. The procedure continues until the average differential extinction and reddening values of all members are less than 0.07 and 0.04, respectively. For NGC\,2158, two iterations were performed, and the effect is obvious, as comparing the original CMD as observed, Fig.~\ref{fig:DR_ngc2158_demo}(d), with that after the correction, Fig.~\ref{fig:DR_ngc2158_demo}(e), for which all evolutionary sequences appear much better defined.  Spatially, a systematic reddening across the cluster is clearly discernible; see Fig.~\ref{fig:DR_ngc2158_demo}(f).

We applied the differential reddening correction to 16 OCs that are significantly influenced by varying degrees of foreground dust. We refrained from implementing this correction to clusters that either already exhibit a very thin and well-defined CMD or are sparse, as this method is based on two key properties: density of members and thickness of MS. Performing the correction for such clusters can obscure the noticeable signs of a split MS and also lose information regarding the position of cluster members on CMDs according to their rotation rates. Moreover, the availability of dust maps is often limited, with many of low resolution, while high-resolution maps require accurate distance measurements for each star. Due to the lack of precise distance measurements for numerous clusters, both types of maps can inadvertently distort CMDs, sometimes misplacing sources from redder parts to bluer parts of the CMDs without accurate corrections being applied.

\begin{figure}
    \centering
\includegraphics[width = 0.45\textwidth]{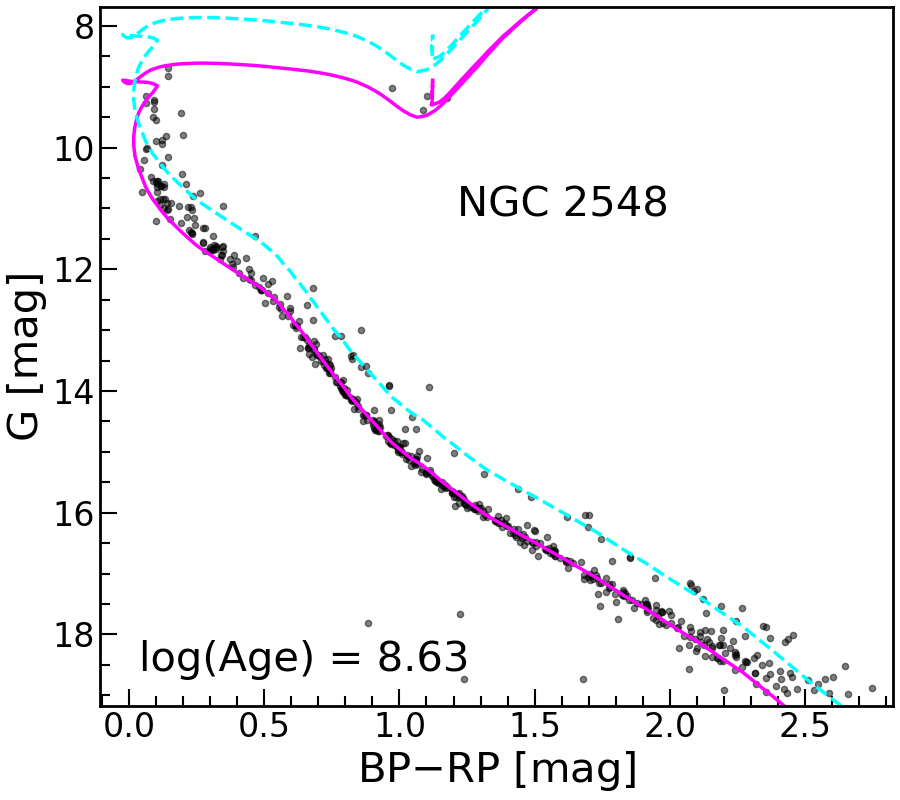}
\includegraphics[width = 0.45\textwidth]{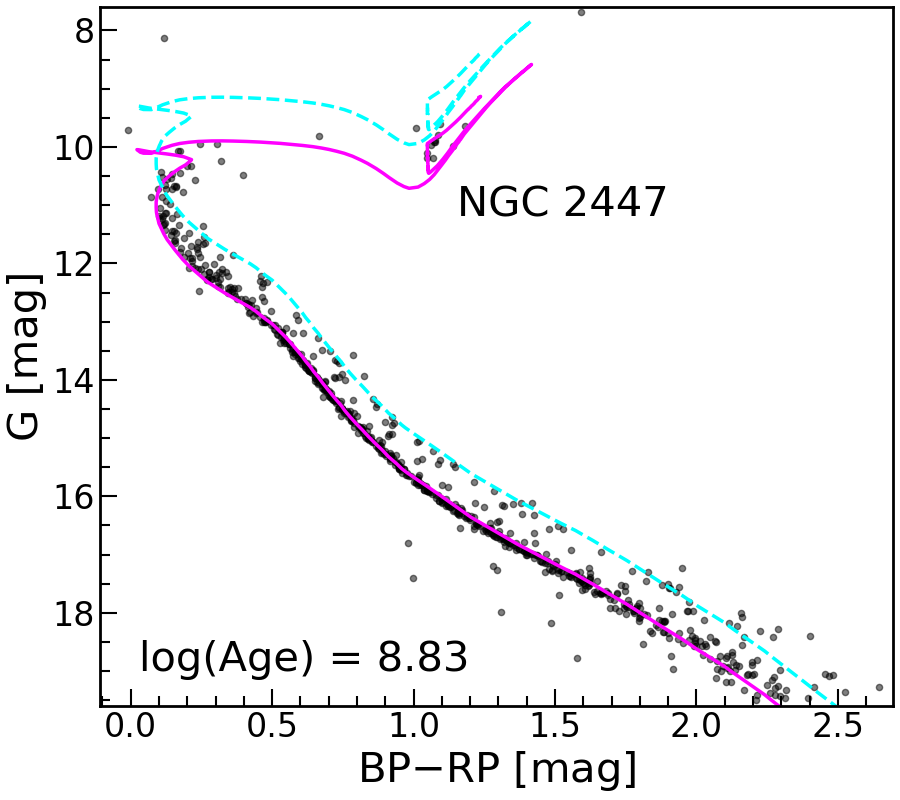}
    \caption{The CMDs of NGC\,2548 and NGC\,2447 with fitted PARSEC isochrones. The fundamental parameters used to fit the isochrones are listed in Table~\ref{tab:fundamental_params}.}
    \label{fig:ISoFit}
\end{figure}

\subsection{Isochrone fitting}

\begin{figure*}
    \centering
    \includegraphics[width=0.9\linewidth]{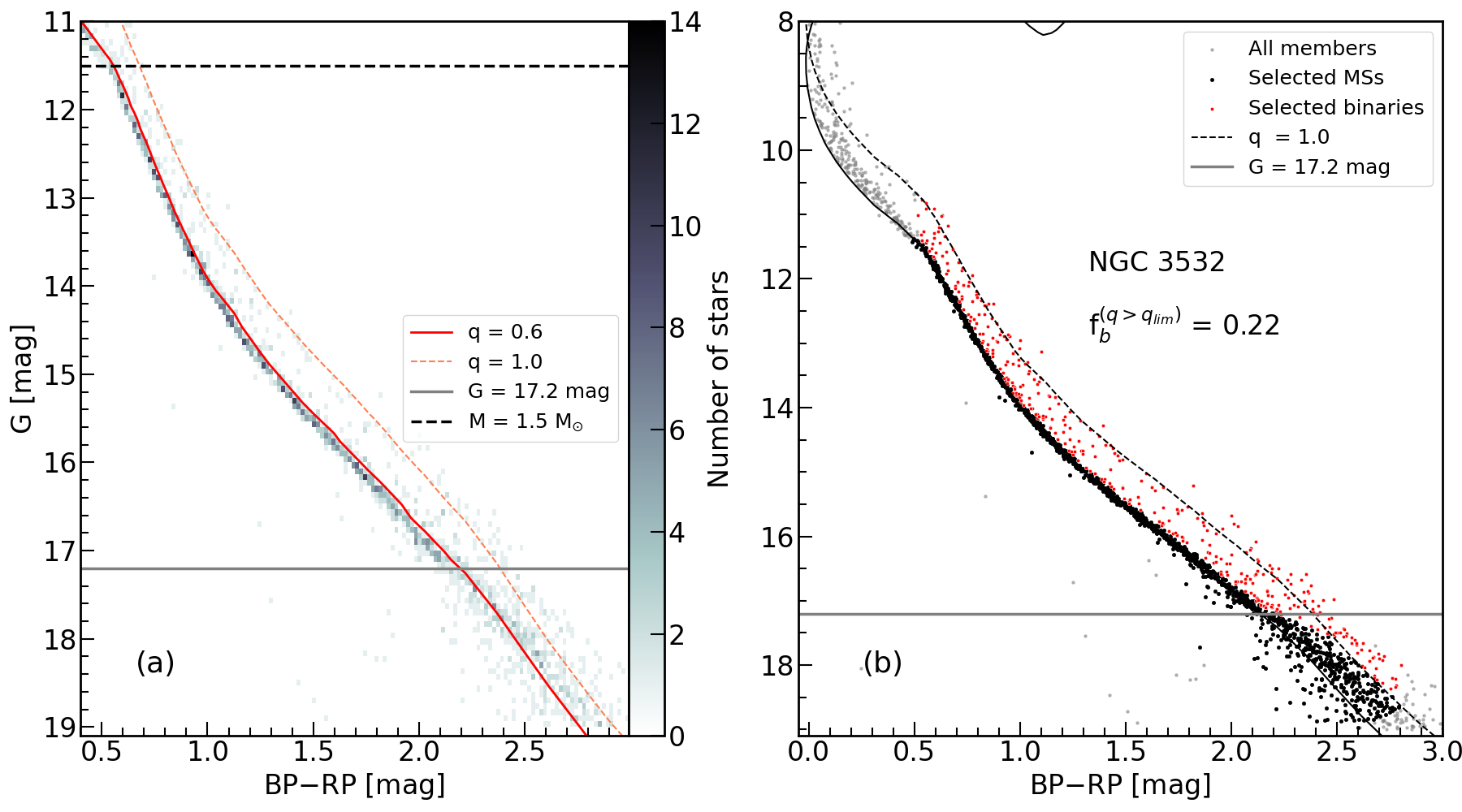}
    \caption{(a): The Hess CMD of NGC\,3532 with a non-rotating isochrone of cluster parameters as listed in \ref{tab:fundamental_params} and $q = 0.6$ (red solid line). The black dashed line is drawn for the single MS mass of 1.5~M$_{\odot}$ for the upper cutoff for the selection of MS and binaries. The grey solid line shows the change in the isochrone from $q = 0.6$--1.0 for the selection of low-mass MS and binaries (see \S\ref{bin} for details). (b): Selected binaries are marked as red, MSs as black, and the rest of the cluster members as grey dots. The red dashed lines in both panels show the non-rotating binary isochrones.}
    \label{fig:binary_sel}
\end{figure*} 

\begin{figure}[!htbp]
    \centering
    \includegraphics[width=0.45\textwidth]{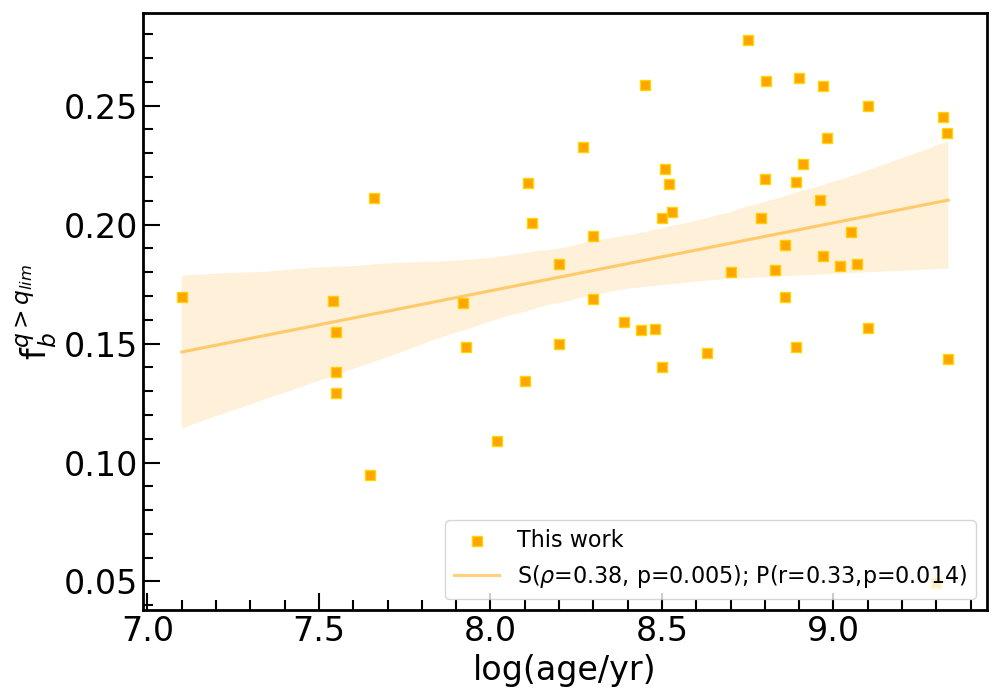}
    \caption{ The correlation of binary fraction with $\log{\rm (age/yr)}$ of the 53 OCs. The solid orange line and spread show the best-fitted relation and $1\sigma$ errors in the correlations. }
    \label{fig:fb_logage}
\end{figure}

To estimate the fundamental parameters of the 53 OCs, we fitted PARSEC V2.0 isochrones \citep{Nguyen2022} to their CMDs. The initial estimates of ages, $A_{\text{V}}$, and distances are taken from \citet{Bhattacharya2022,Rao2023,Cavallo2024}. For metallicities, median $[Fe/H]$ values from \textit{Gaia} DR3 or GES data for cluster members are used. We then adjusted these initial parameters to fit isochrones to the bluer region of a CMD. This approach is motivated by the observation that slow rotators typically occupy the bluer region of a CMD, whereas the fast-rotating members tend to be redder. Fast rotators exhibit positional shifts on CMDs compared to their non-rotating counterparts due to several factors, including gravity darkening, inclination angles, and rotational mixing. 

The equal-mass binaries are located on equal-mass binary isochrones, 0.75 $G$~mags brighter than non-rotating single isochrones, as clearly shown in the CMDs for NGC\,2548 and NGC\,2447 in Fig.~\ref{fig:ISoFit}. This indicates that the isochrone fitting to the blue region of a CMD is appropriate for age estimation. The final parameters thus derived are summarized in Table~\ref{tab:fundamental_params}. 

\subsection{Binary fraction}\label{bin}

We now discuss the selection of unresolved binaries, especially those of high-mass ratios (i.e., a smaller mass difference between component stars). We set a limit on the mass ratio ($q_{\rm lim}$) below which separation of binaries from the populated MS becomes difficult.  The choice of $q_{\rm lim}$ must be brightness dependent; for example, at the faint end, increased photometric errors make binaries almost indistinguishable from the MS. Empirically, $q_{\rm lim}$ ranges from 0.6 to 0.9 for bright stars and from 0.7 to 1.0 for faint stars. As an example, Fig.~\ref{fig:binary_sel} shows the case for NGC\,3532, for which we set $q_{\rm lim} = 0.6$ for brighter stars and $q_{\rm lim} =  1.0$ for those fainter than $G = 17.6$~mag.

The upper mass cutoff for young clusters was determined to be 1.5~M$_{\odot}$. This threshold was established to avoid potential overlap between binaries and fast rotators, which could lead to erroneous classification of fast rotators as binaries. For OCs with MSTO masses around 1.5~M$_{\odot}$ to 1.7~M$_{\odot}$, we used the upper mass cutoff at which the MS begins to bend rightward.  For most clusters, the lower mass cutoff is chosen as the mass of the faintest MS stars. However, in cases where the faint ends of the CMDs are significantly broadened, making it challenging to distinguish binaries, we opted for a slightly higher mass cutoff to ensure accuracy.

We then used the selected binaries and MS stars to estimate the fraction of high mass-ratio binaries using the following equation:
\begin{equation}
    f_b^{q>q_{\rm lim}} = \frac{N_{\rm bin}}{N_{\rm s}+N_{\rm bin}},
\end{equation}
where $N_{\rm bin}$ is the number of selected binaries and $N_{\rm s}$ is the MSs within the same mass range, which, together with the selection parameters of binaries, are presented in Table~\ref{tab:bin_frac}. The binary fractions of the 53 OCs range from 0.10 to 0.30. It is important to note that the binary fraction may also be affected by the effect of uncorrected differential reddening. Nonetheless, the estimated binary fractions for OCs are, by and large, consistent with those reported by \citet{Jadhav2021} and \citet{Jiang2024}.  A broad and moderate positive correlation, as indicated by Pearson and Spearman rank coefficients, is found between binary fraction and age, shown in Fig.~\ref{fig:fb_logage}. A recent study by \citet{Yalyalieva2024} also reported similar results for OCs older than 100~Myr. The increase in binary fraction with age may be due to the fact that while disintegrating, OCs tend to retain binaries because they are relatively massive. 

\begin{table*}[!ht]
    \centering
        \caption{The fundamental parameters of 53 OCs. Here,  Column 1: Cluster name;  Columns 2 and 3: cluster centers;  Columns 4 to 7: fundamental parameters;  Column 8: with or without differential reddening correction;  Columns 9 and 8: the presence of split MS and eMSTO.}
	\label{tab:fundamental_params}
	{\fontsize{7pt}{9pt} \selectfont \begin{tabular}{ccccccccccc}
    \hline
        Cluster & RA & DEC & log(Age) & [M/H] & A$_{\text{v}}$ & distance & Radius & DR corr  & Split MS & eMSTO \\ 
         & (deg) & (deg) &  & (dex) & (mag) & (pc) & (arcmin) &  &  &  \\ \hline
        \\
            
        \multicolumn{11}{c}{\textbf{Class~I OCs}} \\ 
        \\
        
        ASCC 113 & 317.86509176 & 38.644942666 & 8.5 & 0.05 & 0.1 & 560 & 120 & -- & Y & — \\ 
        IC 2602 & 160.530617494 & $-$64.363575791 & 7.55 & 0 & 0.096 & 153 & 200 & -- & Y & — \\ 
        Melotte 22 & 56.7579164568 & 24.1394971043 & 7.93 & 0.05 & 0.13 & 135 & 311 & -- & Y & — \\ 
        NGC 2287 & 101.457699928 & $-$20.725107056 & 8.44 & 0.05 & 0.02 & 728 & 60 & -- & Y & Y \\ 
        NGC 2423 & 114.261077942 & $-$13.924900702 & 9.02 & 0.1 & 0.1 & 905 & 40 & -- & Y & Y \\ 
        NGC 2447 & 116.178197711 & $-$23.872209123 & 8.83 & $-$0.1 & 0.01 & 970 & 50 & -- & Y & Y \\ 
        NGC 2527 & 121.254859227 & $-$28.072913561 & 8.89 & 0 & 0.08 & 620 & 100 & -- & Y & Y \\
        NGC 2539 & 122.688436769 & $-$12.864318333 & 8.86 & 0 & 0.11 & 1245 & 50 & -- & Y & Y \\ 
        NGC 2548 & 123.402350268 & $-$5.709091009 & 8.63 & 0.05 & 0.05 & 750 & 100 & -- & Y & Y \\ 
        NGC 2632 & 129.905357614 & 19.8309828888 & 8.86 & 0.1 & 0 & 175 & 234 & -- & Y & — \\ 
        NGC 3114 & 150.551954036 & $-$60.128326517 & 8.3 & 0 & 0.21 & 973 & 80 & -- & Y & Y \\ 
        NGC 3532 & 166.359666875 & $-$58.69382408 & 8.52 & 0.12 & 0.07 & 490 & 120 & -- & Y & Y \\ 
        NGC 6811 & 294.373513672 & 46.3761779755 & 8.98 & 0 & 0.12 & 1100 & 40 & -- & Y & Y \\ 
        Trumpler 10 & 132.151655822 & $-$42.67402699 & 7.55 & 0 & 0.15 & 450 & 104 & -- & Y & — \\

        \\

        \multicolumn{11}{c}{\textbf{Class~II OCs}} \\ 
        \\
        
        Collinder 463 & 26.968742972 & 71.771404496 & 8.45 & 0 & 0.64 & 805 & 100 & $\checkmark$ & — & Y \\ 
        NGC 1912 & 82.1072291355 & 35.8565688065 & 8.5 & $-$0.05 & 0.71 & 1148 & 50 & $\checkmark$ & — & Y \\ 
        NGC 2099 & 88.1117239813 & 32.5715927702 & 8.79 & $-$0.02 & 0.54 & 1400 & 50 & $\checkmark$ & Partial & Y \\ 
        NGC 2168 & 92.2843679614 & 24.3176256903 & 8.11 & $-$0.1 & 0.5 & 867 & 60 & $\checkmark$ & — & Y \\ 
        NGC 1039 & 40.6251036262 & 42.809115747 & 8.12 & 0.1 & 0.18 & 500 & 80 & -- & Y & — \\ 
        NGC 2301 & 102.956707892 & 0.43059767662 & 8.2 & 0 & 0.19 & 847 & 70 & -- & Y & — \\ 
        NGC 2360 & 109.491446357 & $-$15.635742998 & 9.07 & $-$0.05 & 0.23 & 1070 & 50 & -- & Y & Y \\ 
        NGC 2422 & 114.111318599 & $-$14.479119175 & 8.1 & 0.1 & 0.22 & 488 & 80 & -- & Y & — \\ 
        NGC 2437 & 115.436020953 & $-$14.848464533 & 8.48 & $-$0.05 & 0.3 & 1600 & 50 & $\checkmark$ & Y & Y \\ 
        NGC 2516 & 119.414278926 & $-$60.700156517 & 7.92 & 0 & 0.41 & 420 & 141 & -- & Y & Y \\ 
        NGC 3293 & 158.981576669 & $-$58.236385402 & 7.1 & 0 & 0.76 & 2489 & 10 & -- & — & $\checkmark$ \\ 
        NGC 3766 & 174.069592833 & $-$61.613921181 & 7.54 & $-$0.05 & 0.65 & 1845 & 12.5 & -- & — & Y \\ 
        NGC 5822 & 226.156036364 & $-$54.364302045 & 8.97 & 0 & 0.38 & 830 & 52 & -- & Y & Y \\ 
        NGC 6067 & 243.268834982 & $-$54.235032862 & 8.2 & 0.15 & 0.9 & 2000 & 14 & $\checkmark$ & — & Y \\ 
        NGC 6281 & 256.169277122 & $-$37.93739437 & 8.51 & $-$0.02 & 0.43 & 505 & 65 & -- & Y & Y \\ 
        NGC 6633 & 276.80856402 & 6.63888959683 & 8.75 & $-$0.05 & 0.53 & 400 & 160 & -- & Y & Y \\ 
        NGC 6649 & 278.360075323 & $-$10.392935242 & 7.65 & 0.15 & 3.85 & 1655 & 14 & $\checkmark$ & — & Y \\ 
        NGC 6705 & 282.76271461 & $-$6.2767033473 & 8.53 & 0.01 & 0.84 & 1835 & 12 & $\checkmark$ & Y & Y \\ 
        NGC 6940 & 308.634997009 & 28.2549887733 & 8.96 & 0.13 & 0.36 & 1055 & 41 & -- & Y & Y \\ 
        NGC 7209 & 331.304701622 & 46.4971641516 & 8.8 & $-$0.05 & 0.35 & 1090 & 40 & -- & Y & Y \\ 
        \\ \hline
    \end{tabular}}
\end{table*}

\begin{table*}[!ht]
    \centering
	{\fontsize{7pt}{9pt} \selectfont \begin{tabular}{ccccccccccc}
    \hline
        Cluster & RA & DEC & log(Age) & [M/H] & A$_{\text{v}}$ & distance & Radius & DR corr & Split MS & eMSTO \\ 
         & (deg) & (deg) &  & (dex) & (mag) & (pc) & (arcmin) &  &  &  \\ \hline
        \\
        
        \multicolumn{11}{c}{\textbf{Class~III OCs}} \\ 
        \\

        Alessi 1 & 13.2985447025 & 49.4849173963 & 8.97 & 0.15 & 0.09 & 720 & 40 & -- & — & — \\ 
        Alessi 6 & 220.052499214 & $-$66.155204351 & 8.803 & $-$0.1 & 0.69 & 813.58 & 60 & -- & Partial & — \\ 
        IC 4665 & 266.39093584 & 5.67125277893 & 7.66 & 0.05 & 0.44 & 345 & 60 & -- & Y & — \\ 
        NGC 1901 & 79.4431376721 & $-$68.241192712 & 8.9 & $-$0.08 & 0.1 & 420 & 55 & -- & Partial & — \\ 
        NGC 2451B & 115.697552875 & $-$37.40910812 & 7.55 & 0 & 0.18 & 360 & 120 & -- & Y & — \\ 
        NGC 2482 & 118.823582157 & $-$24.254683815 & 8.7 & 0 & 0.12 & 1284 & 40 & -- & Y & Y \\ 
        NGC 5460 & 211.730485371 & $-$48.252900505 & 8.27 & 0 & 0.32 & 725 & 60 & -- & Partial & — \\ 
        NGC 6005 & 238.963548474 & $-$57.435006558 & 9.05 & 0.19 & 1.16 & 2240 & 13 & $\checkmark$ & Partial & Y \\ 
        NGC 6025 & 240.669738757 & $-$60.43198321 & 8.3 & 0.08 & 0.39 & 765 & 75 & -- & — & — \\ 
        NGC 6475 & 268.570348069 & $-$34.657293055 & 8.39 & 0.05 & 0.22 & 275 & 163 & -- & Y & Y \\ 
        NGC 6802 & 292.664556366 & 20.2428940639 & 8.91 & 0.1 & 2.26 & 2070 & 10 & $\checkmark$ & — & Y \\ 
        Pismis 15 & 143.702965198 & $-$48.055263898 & 9.1 & 0.07 & 1.62 & 2023 & 9 & $\checkmark$ & — & — \\ 
        Roslund 6 & 307.542634889 & 39.8525344268 & 8.02 & 0.05 & 0.04 & 345 & 60 & -- & Y & — \\ 
        Ruprecht 134 & 268.195415083 & $-$29.530459014 & 9.1 & 0.25 & 1.2 & 2040 & 9 & $\checkmark$ & Y & Y \\ 
        Trumpler 23 & 240.202857418 & $-$53.538932574 & 8.89 & 0.15 & 1.78 & 2060 & 13 & $\checkmark$ & Y & Y \\ 
        \\
        \multicolumn{11}{c}{\textbf{Class~IV OCs}} \\ 
        \\
        Trumpler 20 & 189.898191731 & $-$60.640198181 & 9.3 & 0.15 & 0.95 & 2800 & 13 & $\checkmark$ & — & Y \\ 
        NGC 2141 & 90.7471364739 & 10.4556171864 & 9.33 & $-$0.05 & 0.8 & 4200 & 12 & $\checkmark$ & — & — \\ 
        NGC 2158 & 91.8606057492 & 24.098068029 & 9.32 & $-$0.15 & 1.11 & 4000 & 10 & $\checkmark$ & — & — \\ 
        NGC 2420 & 114.612635122 & 21.573031362 & 9.335 & $-$0.15 & 0.05 & 2500 & 15 & -- & — & — \\
\\ \hline
    \end{tabular}}
\end{table*}

\section{Results \& Discussion} \label{sec:analysis}

OCs span a wide range of ages, masses, metallicities, and Galactic environments, and their stellar populations are continuously shaped by both internal dynamical evolution and external interactions with the surrounding interstellar medium. These factors collectively influence the morphology of the upper MS and MSTO and the distribution of fast and slow rotators.
Therefore, in this section, we analyze the morphologies and rotational distribution of the 53 OCs and systematically investigate the effects of extinction, age, mass loss, and binary fraction on MSTO morphology and the slow-rotator population.

\subsection{Clusters classification}\label{morpho}

\begin{figure*}[htbp]
    \centering
    \includegraphics[width=0.9\textwidth]  
       {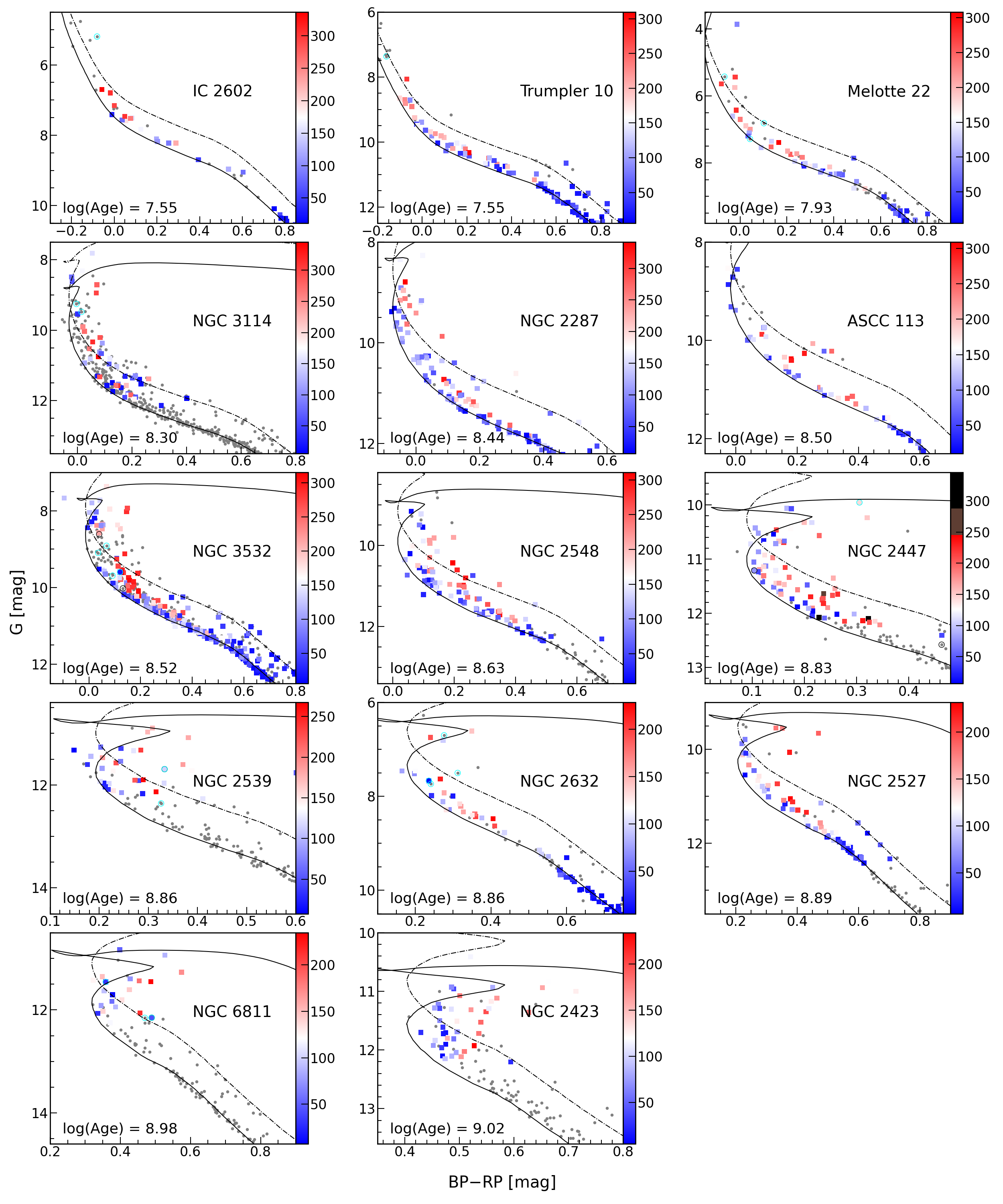}
    \caption{CMDs of Class~I OCs showing a clear bimodal $v \sin i$ distribution in the upper MS and MSTO regions. Cluster members are represented by grey symbols, and those with available $v \sin i$ measurements by squares, color-coded by rotation level according to the associated colorbars. The black solid and dashed lines depict the fitted isochrones for single stars and for equal-mass binaries, respectively, based on the fundamental parameters listed in Table~\ref{tab:fundamental_params}.  
        }
    \label{fig:split_ms}
\end{figure*}

\begin{figure*}[htbp]
    \centering
    \includegraphics[width=0.9\linewidth]
       {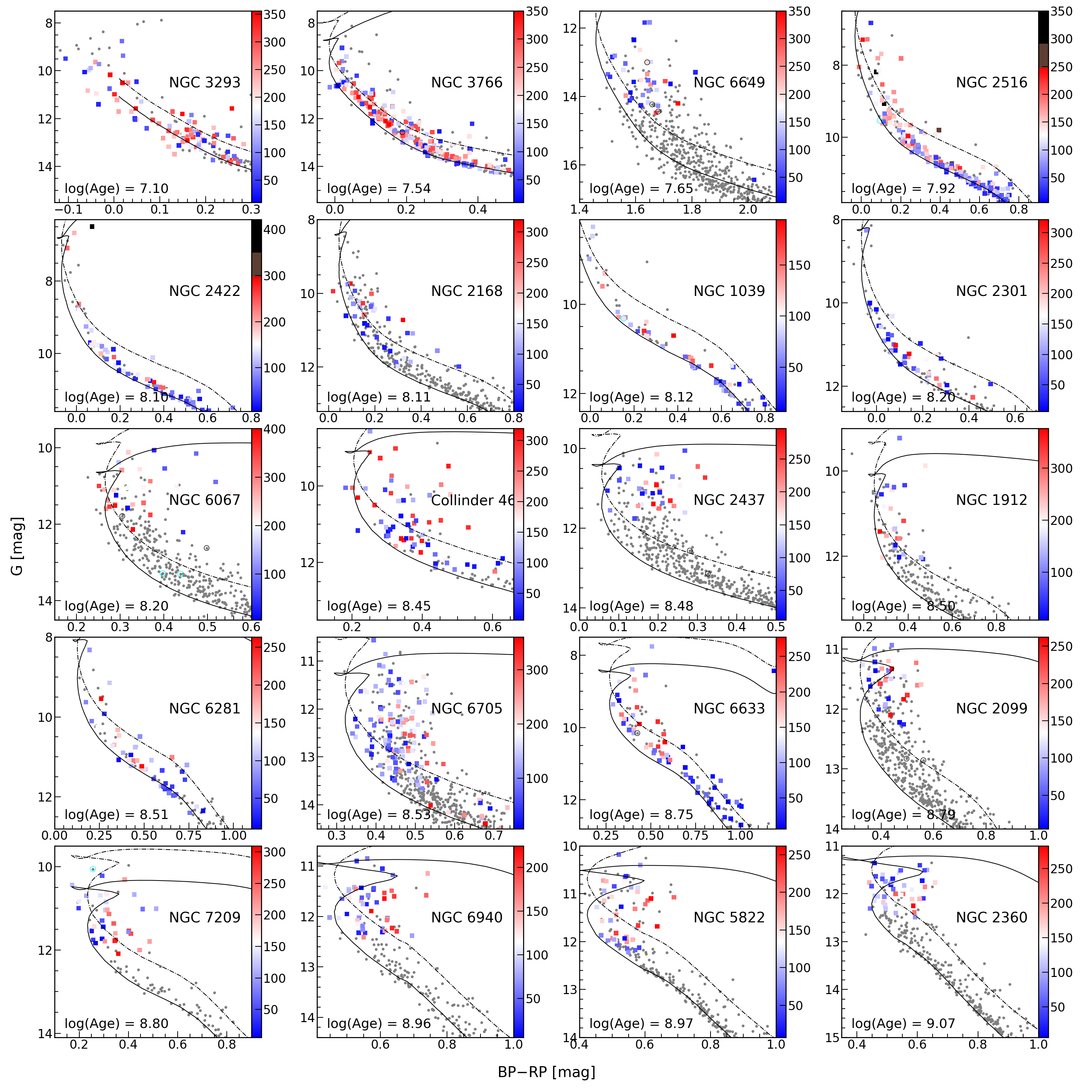}
    \caption{The same as in Fig.~\ref{fig:split_ms} but for the Class~II OCs, which have minimal to moderate distinction between fast and slow rotators.}
    \label{fig:eMSTO}
\end{figure*}

\begin{figure*}[htbp]
    \centering
    \includegraphics[width=0.9\textwidth]
      {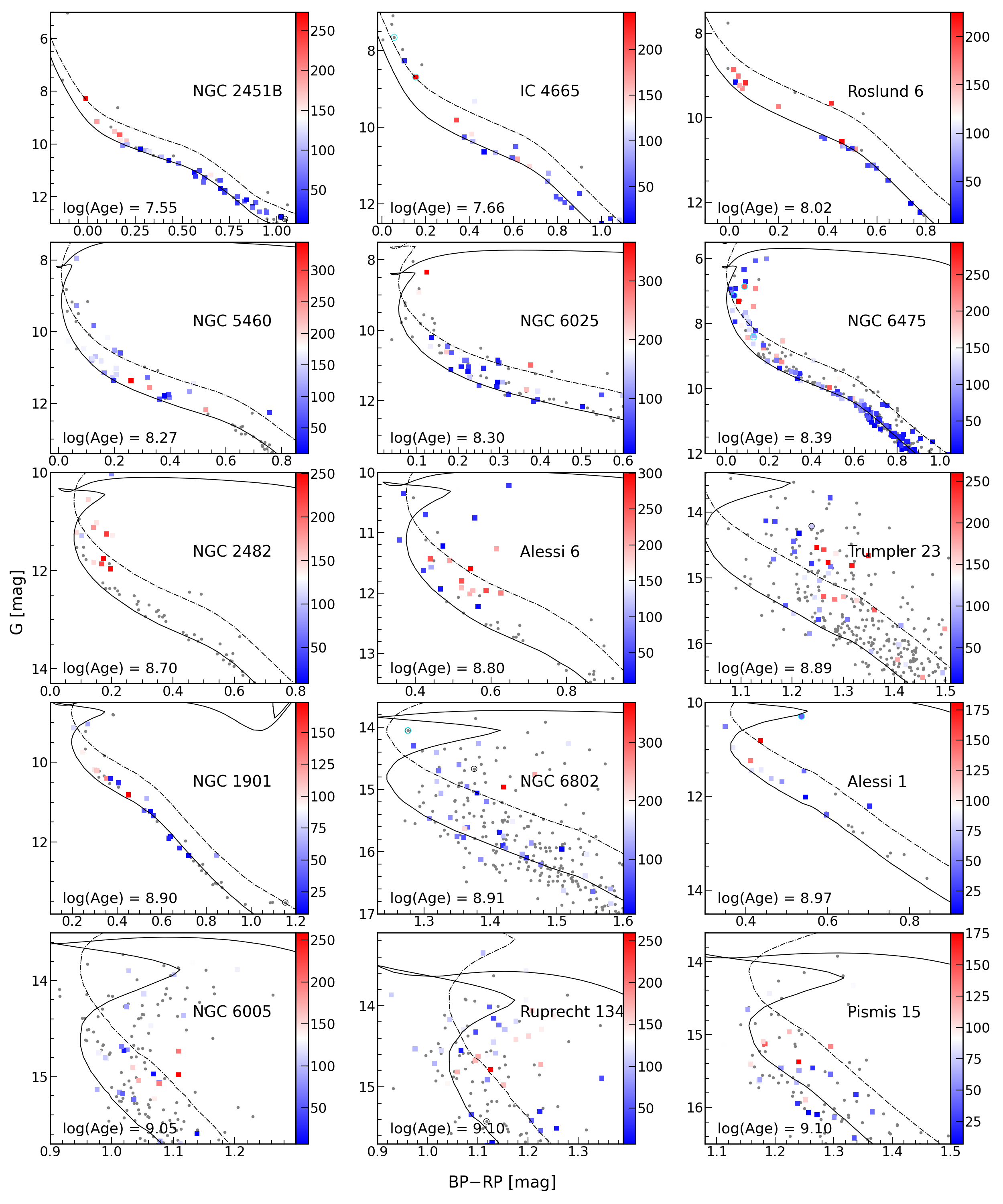}   
    \caption{The same as in Fig.~\ref{fig:split_ms} but for Class~III OCs}.
    \label{fig:weird_ocs}
\end{figure*}

\begin{figure}[htbp]
    \centering
    \includegraphics[width=0.98\linewidth] 
      {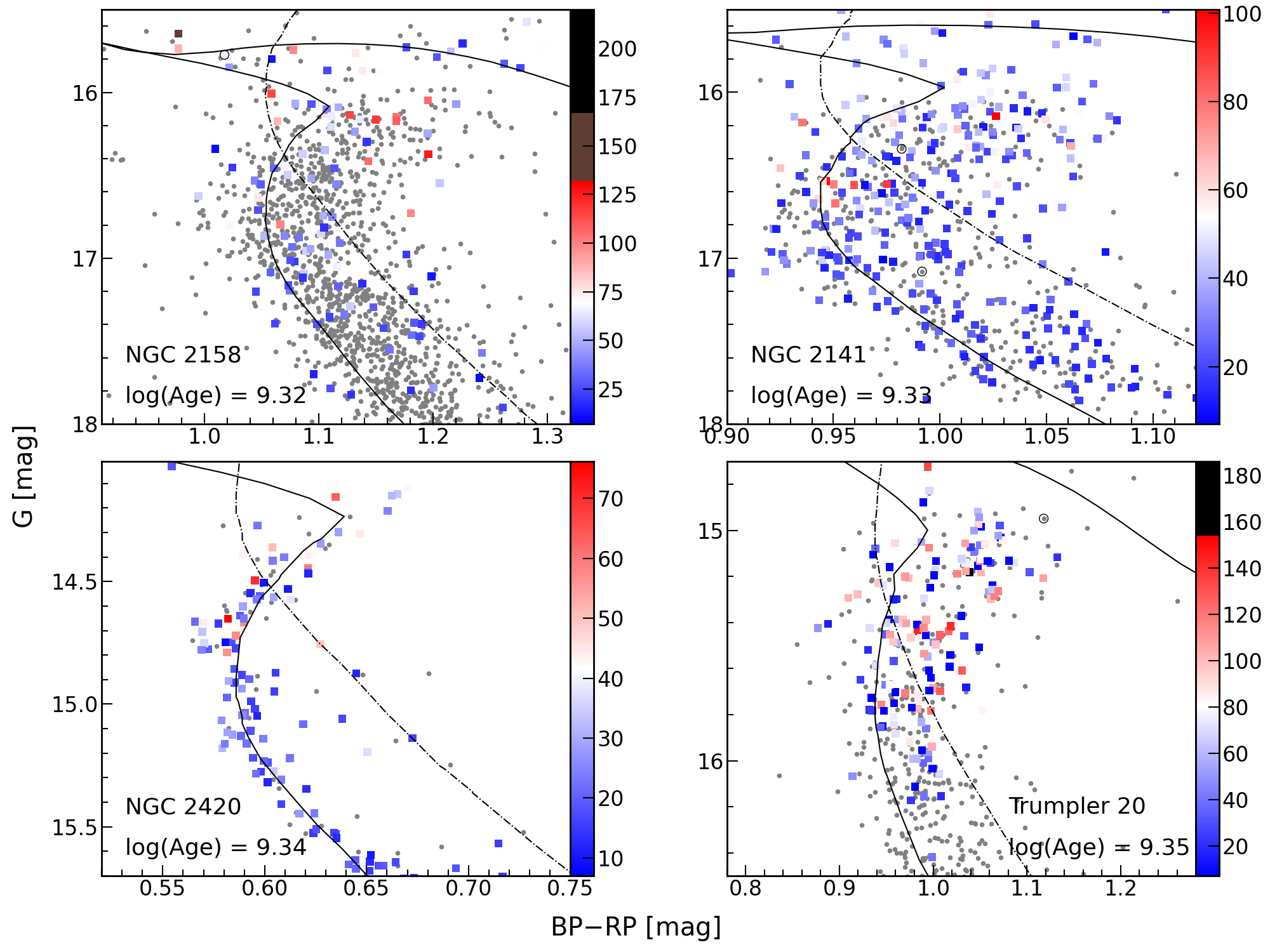}
    \caption{The same as in Fig.~\ref{fig:split_ms} but for Class~IV OCs that} older than 2~Gyr.
    \label{fig:old_ocs}
\end{figure}

To explore the morphology of the upper MS and MSTO and possible links to stellar rotation or other cluster properties, we classified the OCs into four distinct groups, shown in Fig.~\ref{fig:split_ms}, Fig.~\ref{fig:eMSTO}, Fig.~\ref{fig:weird_ocs}, and Fig.~\ref{fig:old_ocs}.

\begin{enumerate}

\item \textit{Class~I:} 14 OCs are classified into this category based on their minimal interstellar extinction. Across these OCs, as shown in Fig.~\ref{fig:split_ms}, the majority of rapidly rotating stars occupy redder sides of MS in CMDs. At the same time, slow rotators preferentially populate bluer sides. The redder sequences still have a few slow rotators due to unresolved binaries, as high mass-ratio binaries overlap with the locus of fast rotators \citep{Rao2025arXiv251205458R}. Since these OCs are minimally affected by interstellar extinction, this enables the detection of the intrinsic locations of fast and slow rotators. These clusters are thus called “golden” samples, which should be used for investigating intrinsic MSTO morphology, the origin of fast and slow rotators, and for calibrating rotating stellar evolutionary models, and should be prioritized in future comparisons with theoretical predictions. We have nine OCs in common with \citet{Cordoni2024}. The remaining five OCs are analyzed here for the first time, using $v \sin i$ measurements from \textit{Gaia} DR3 and GES, thereby extending the golden sample beyond what has been previously available in the literature. Interestingly, NGC\,2287 clearly shows bifurcated MS in the upper MS within $G = 10.5$--11.5~mag, which is also reported in \citet{Sun2019}.  

\item \textit{Class~II:} This class contains 20 OCs exhibiting minimal to moderate photometric distinctions between fast and slow rotators, as shown in Fig.~\ref{fig:eMSTO}. These clusters consistently display apparent eMSTO features. In particular, they are moderately to highly influenced by interstellar extinction. Even after differential reddening correction, residual extinction effects may still displace stars in the CMD, potentially causing misidentification of fast rotators as slow rotators or vice versa. Therefore, direct interpretation of the formation mechanisms of their fast and slow rotators based on CMD and spatial positions should be avoided. However, these clusters contain large numbers of MSTO stars and therefore are the best samples for studying the origin of extremely fast rotators and for exploring the role of different spin-down mechanisms. Six of these OCs are in common with \citep{Cordoni2024}. Two of these OCs have also been subjected to detailed studies for the formation of slow and fast rotators, including NGC\,6067 \citep{Maurya2025ApJ...989..123M} and NGC\,6940 \citep{Panthi2024}.

\item \textit{Class~III:} A total of 14 OCs are classified into this group, containing very few stars in their upper MS or overall broad CMDs to exhibit eMS/eMSTO features, as shown in Fig.~\ref{fig:weird_ocs}.  Similar to Class~II clusters, they do not show clear photometric separation between fast and slow rotators. The small number of MSTO stars in these clusters is not necessarily indicative of an absence of intrinsic rotation distribution, but rather reflects the consequences of significant mass loss over the cluster lifetime.

\item \textit{Class~IV:} Four OCs, namely NGC\,2141, NGC\,2158, NGC\,2420, and Trumpler\,5 belong to this group. Their MSTO corresponds to the stellar mass between 1.3~M$_{\odot}$ and 1.6~M$_{\odot}$, the threshold for magnetic braking on the MS. These clusters have comparable ages ($9.3 < \log({\rm (age/yr}) < 9.35$) but vary in metallicity, making them ideal targets for investigating the onset of magnetic braking as a function of cluster metallicity. 

\end{enumerate}

\subsection{MSTO width vs age correlation}\label{Av}

We now consider the effect of extinction, cluster age, and masses that may affect the morphology and width of the MSTO.

\begin{figure*}
    \centering
\includegraphics[width=0.4\textwidth]{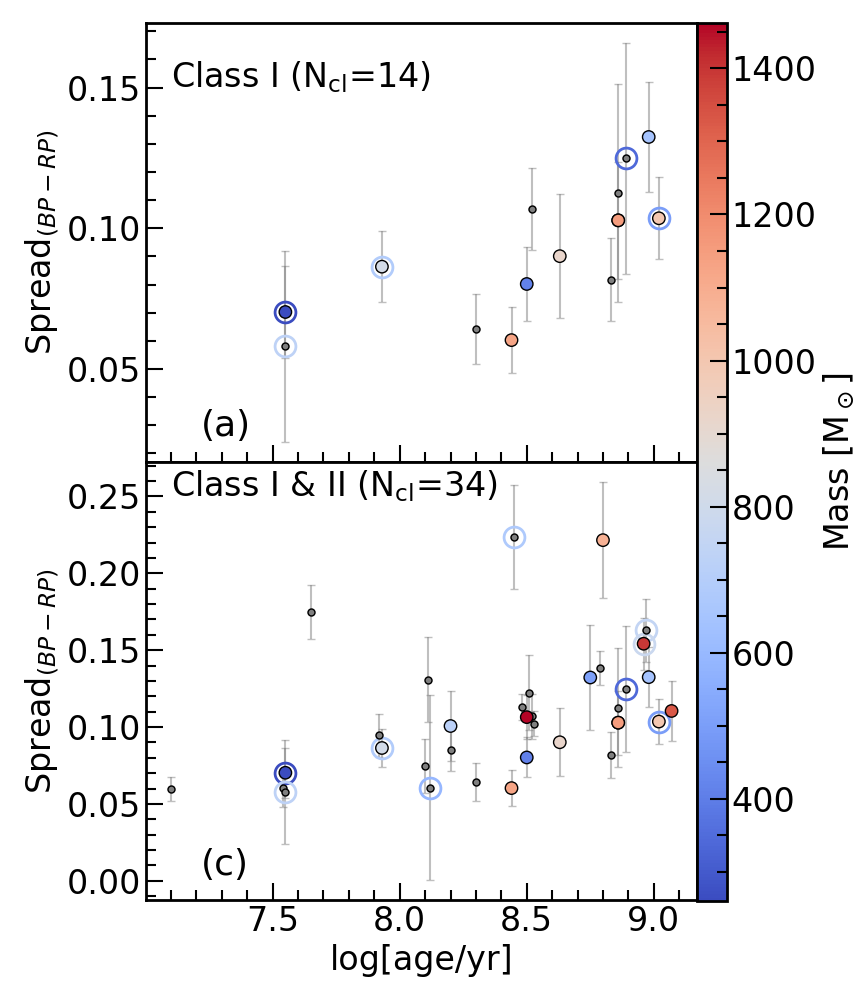}       
\includegraphics[width=0.4\textwidth]{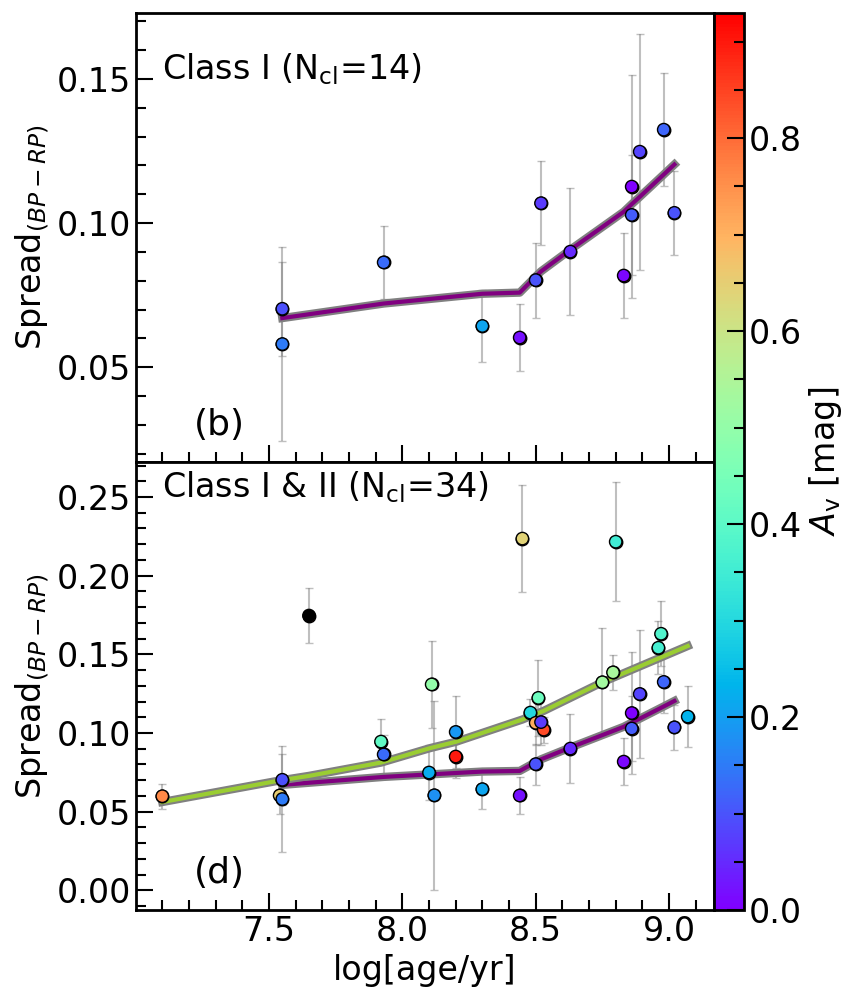}
    \caption{A correlation of the color spread of MSTO stars with cluster ages. (a) and (b) are for Class I OCs, and (c) and (d) are for Class I \& Class II OCs. (a) and (c): Colorbar represents masses of OCs taken from \citet{Almeida2023} and \citet{Bhattacharya2022}, respectively.  (b) and (d): Color shows $A_{\rm v}$ values as listed in Table~\ref{tab:fundamental_params}. The black datapoint in (d) is NGC\,6649 OC with the highest $A_{\rm v}$ value of $\sim 3.85$~mag. The purple and green trend lines represent the fitted LOWESS models for Class I OCs and Class I \& II OCs, respectively.}
    \label{fig:fwhm}
\end{figure*}

To investigate the effect of extinction and mass on the relationship between MSTO broadness and age, we selected MSTO stars with $G = 0.2$--6~mag, based on the ages of the OCs, and quantified MSTO width using the color spread. For this, we first computed the normalized color values ($\Delta BP-RP$~mag) of the selected MSTO stars using fitted isochrones as a reference line. The color spread is parameterized from the 84th to the 16th percentiles of the Weibull distribution. The uncertainties in the MSTO width are estimated using bootstrap sampling, drawing 1000 samples from the selected MSTO and calculating the quantile width for each subset. The standard deviation of these quantile widths is used as the error linked to the estimated color spreads. The number of MSTO stars and estimated color spreads and associated errors are listed in Table~\ref{tab:bin_frac}. Fig~\ref{fig:fwhm} shows the correlation of the color spread of MSTO stars of the 53 OCs in our study with their age, with mass or extinction additionally color-coded. OCs' masses are obtained from \citet{Bhattacharya2022} and \citet{Almeida2023}, which include 15 OCs common to \citet{Bhattacharya2022} and 22 OCs common to \citet{Almeida2023}.  

We observed that the MSTO width tends to increase with cluster age as previously observed by \citet{Niederhofer2015MNRAS.453.2070N}, \citet{Cordoni2018}, and \citep{Cordoni2024}. 
No apparent effect of cluster masses on the correlation is observed, although this is limited by the different mass estimation methods used in the literature \citep{Bhattacharya2022, Almeida2023}. We observed that OCs having higher $A_{\rm v}$ values consistently have a larger color spread and OCs having smaller $A_{\rm v}$ values consistently have a smaller color spread, as shown in \ref{fig:fwhm}(b) and \ref{fig:fwhm}(d), which naturally broadens the overall correlation. In this work, we adopted a single extinction value per cluster for isochrone fitting to characterize the cluster's global extinction. Although we applied differential reddening corrections to nearly half of our sample, these corrections do not fully account for small-scale variations in extinction across individual clusters. As a result, high or spatially variable extinction introduces photometric shifts that artificially inflate the apparent MSTO width. Therefore, higher extinction introduces a systematic offset in the measured MSTO color spread that must be accounted for when comparing observations directly with predictions from rotating stellar models. Nonetheless, the correlation remains evident in both high- and low-extinction OCs, underscoring the robustness of the correlation. It confirms that the correlation is physical, given that our sample spans a wide range of Galactic environments. 

\subsection{Open clusters near the magnetic braking limit }

\begin{figure*}
    \centering
    \includegraphics[width=0.8\textwidth]{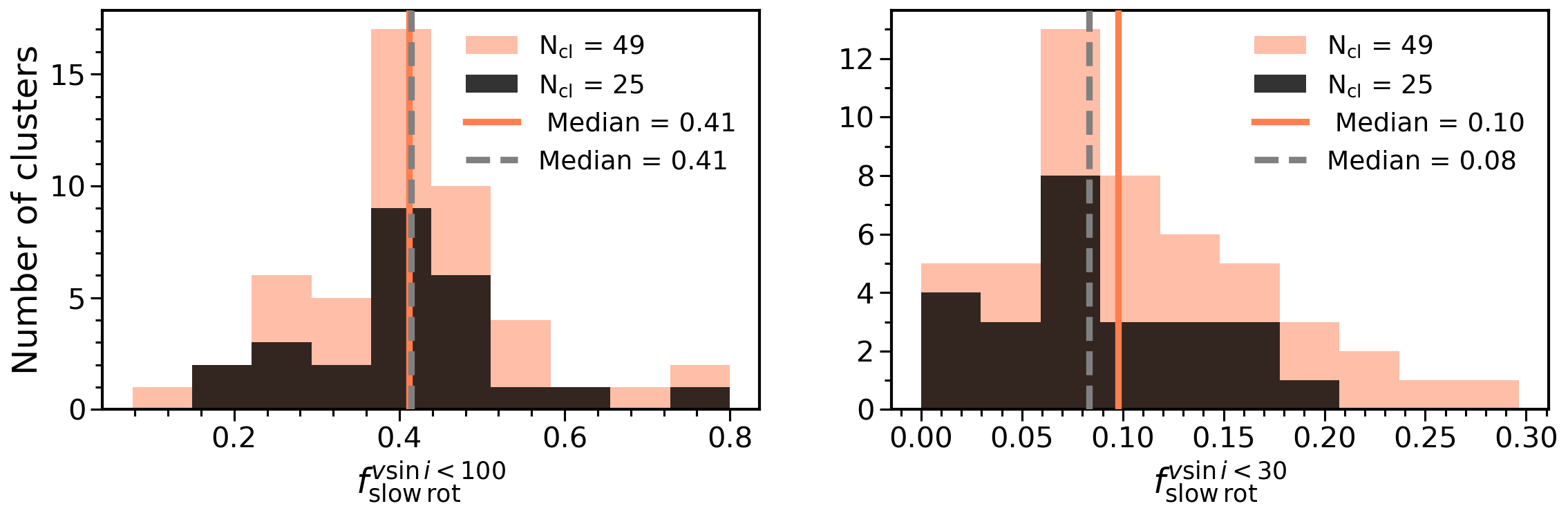}
    \includegraphics[width=0.45\linewidth]{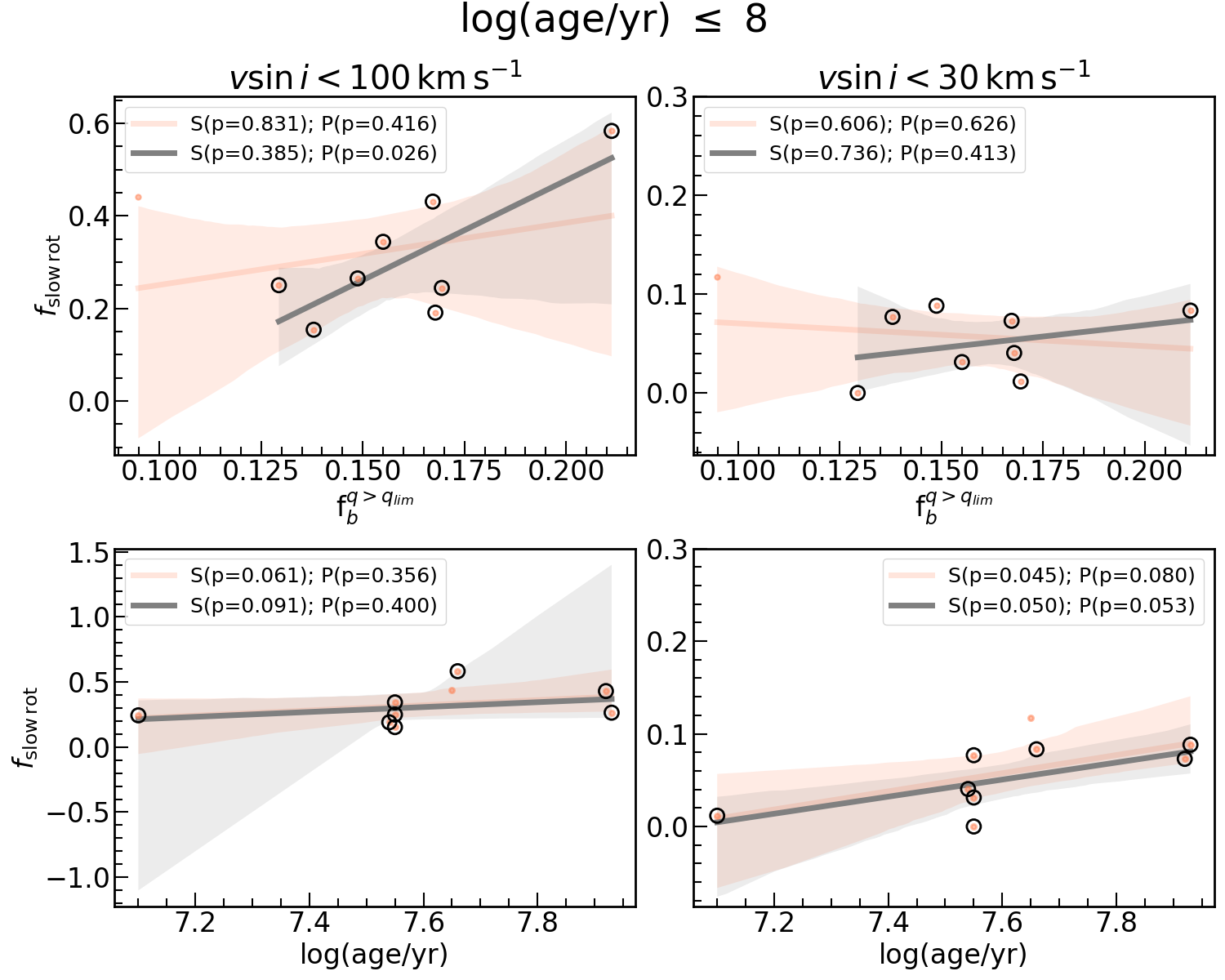}
        \includegraphics[width=0.45\linewidth]{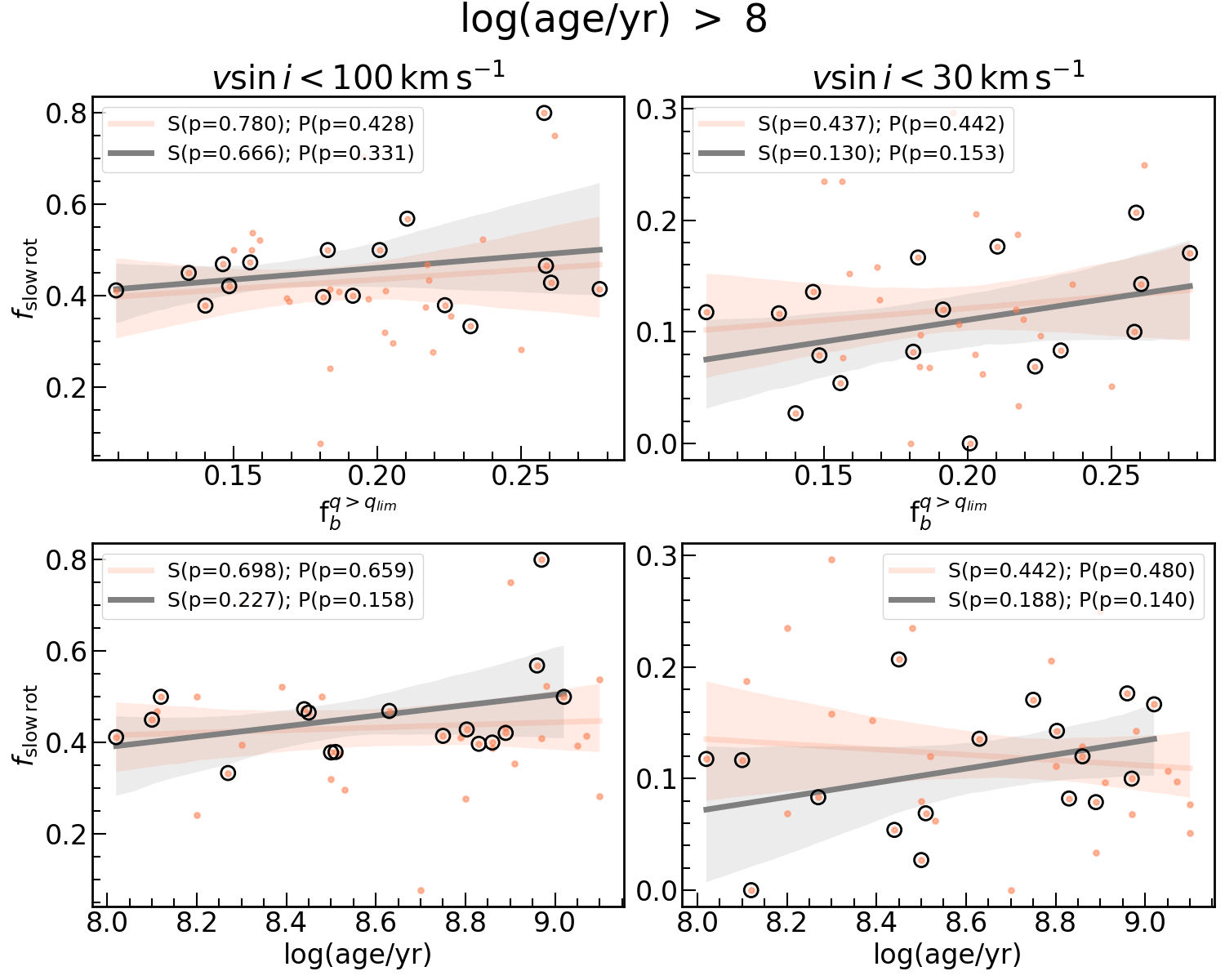}
    \caption{Distribution of fractions of slow rotators (upper panels) and correlations of fractions of slow rotators with high-mass-ratio binary fractions (middle panels) and $\log{\rm (age/yr)}$ (lower panels) and for different age groups and within different $v \sin i$ limits. The coral color represents all 49 Class I, II, and III OCs, whereas the black color represents 25 Class I, II, and III OCs with at least 50\% MSTO stars for which $v \sin i$ measurements are available.}
    \label{fig:slow_rot}
\end{figure*}

Four OCs, NGC\,2141, NGC\,2420, NGC\,2158, and Trumpler\,20, of Class~IV are older than $\log{\rm (age/yr)} = 9.3$ or about 2~Gyr, which seems to mark the empirical upper age limit for the eMSTO phenomenon. As seen in Fig.~\ref{fig:old_ocs}, the upper MS of NGC\,2420 appears well defined with no signs of eMSTO, whereas the other three clusters have relatively broad MSs. Trumpler\,20 hosts a large number of fast rotators with $v \sin i > 100$~km~s$^{-1}$, while NGC\,2141 and NGC\,2420 contain predominantly slow rotators with $v \sin i < 70$~km~s$^{-1}$. 
NGC\,2158, on the other hand, has a handful of sources with $v \sin i$ measurements to make any firm conclusion about the overall rotational distribution.

Although the isochrone ages of these four clusters are comparable, their metallicities differ. Trumpler\,20 is metal-rich ($[Fe/H] = +0.15$~dex), NGC\,2141, NGC\,2158 and NGC\,2420 are sub-solar with values of $[Fe/H]$ ranging from $-0.15$~dex to $-0.05$~dex with turnoff masses $\sim 1.5$~M$_\odot$. Whereas for Trumpler\,20, the turnoff mass is $\sim 1.6$--1.7~M$_\odot$. Although Trumpler\,20 is comparatively metal-rich, it is slightly younger; therefore, it has not yet experienced significant magnetic braking. This old clusters sample suggests that clusters of sub-solar metallicity experience magnetic braking around $\sim 1.5$~M$_\odot$. The range of metallicity for these four clusters is too small to see the complete effect of metallicity on the exact mass limit of magnetic braking. Our results are consistent with those of \citep{Amard2020ApJ...889..108A, See2024MNRAS.533.1290S} that demonstrated that metal-rich stars experience magnetic braking at relatively early ages.

\begin{figure*}
    \centering
    \includegraphics[width=0.95\textwidth]{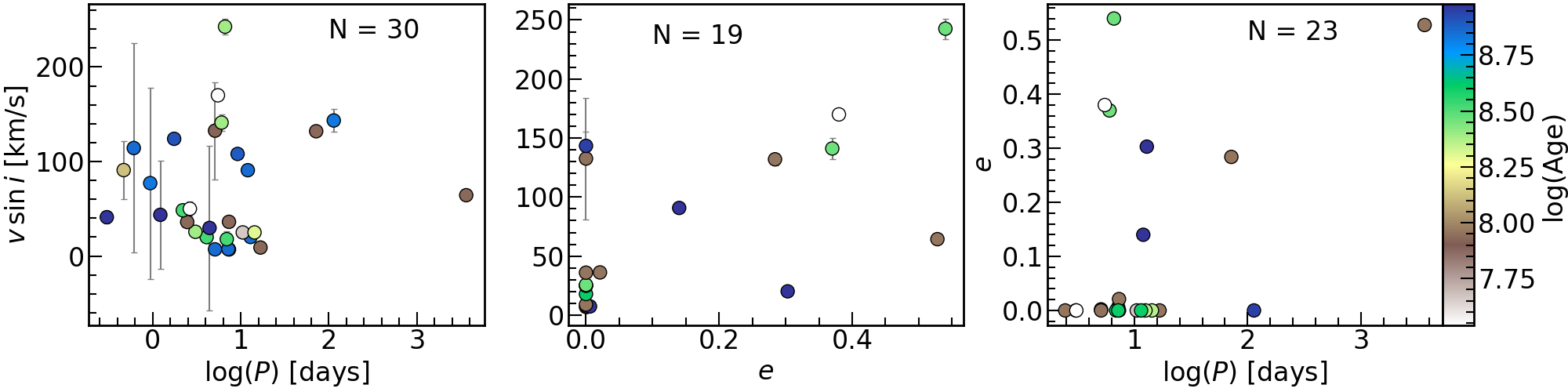}
    \caption{Correlation of orbital parameters of MSTO eclipsing binaries and spectroscopic binaries with their $v  \sin i$ values. }
    \label{fig:bin_var}
\end{figure*}

\subsection{Fraction of slow rotators}\label{slow_rot}

The distinction between the fast- and slow-rotating populations is biased in the sense that a fast rotator can have an observed smaller $v \sin i$ because of the projection, but a star with a large $v \sin i$ must intrinsically be a fast rotator.  

Here we estimate the fraction of slow rotators ($f_{\rm slow \, rot}$) among MSTO stars having masses greater than 1.5~M$_{\odot}$ in Class I, II, and III OCs. As described in \S\ref{Av}, we excluded binaries from the MSTO samples that are clearly identifiable along the fitted equal-mass binary isochrones. For this analysis, we evaluated the fraction of slow rotators in two specific velocity ranges: $v  \sin i < 100$~km~s$^{-1}$ ($f_{\rm slow\,  rot}^{v \sin i<100}$) and $v \sin i < 30$~km~s$^{-1}$ ($f_{\rm slow\, rot}^{v \sin i<30}$). The observed $f_{\rm slow\,  rot}$ values are upper limits on the intrinsic slow-rotator fraction, since a fast rotator viewed at low inclination can mimic a slow rotator. The $f_{\rm slow\,  rot}$ values are listed in Table~\ref{tab:bin_frac}.

Fig~\ref{fig:slow_rot} (upper panels) shows the distribution of $f_{\rm slow\,  rot}^{v \sin i<100}$ and $f_{\rm slow\,  rot}^{v \sin i<300}$. The median value of $f_{\rm slow\, rot}^{v \sin i<100}$ distribution is $\approx 0.41$.  For statistically reliable results, we overplotted 25 Class I, II, and III OCs that have at least 50\% MSTO stars with $v \sin i$ measurements. The smaller and higher ends of the $f_{\rm slow\, rot}^{v \sin i<100}$ distribution are occupied by NGC\,2451B ($f_{\rm slow\, rot}^{v \sin i<100} = 0.15$) and Alessi\,1 ($f_{\rm slow\, rot}^{v \sin i<100} = 0.8$), respectively. Both belong to Class~III OCs and are currently experiencing mass-loss as observed by their tidal tail features \citep{Bhattacharya2022, Tarricq2022A&A...659A..59T}. This indicates that their hostile cluster environment might have altered the intrinsic rotational distribution of their MSTO stars. The small number statistics and inclination could also affect the estimated fractions, but may also reflect environment-dependent spin-down mechanisms. 

The distribution of $f_{\rm slow\, rot}^{v \sin i<30}$ is quite broad, with the median value $\approx 0.08$. The sharp contrast between median values of $f_{\rm slow\, rot}^{v \sin i<30} $ (0.08) and $f_{\rm slow\, rot}^{v \sin i<100}$ (0.41) indicates that while a substantial fraction of MSTO stars rotate moderately slowly, very few reach the slow rotation limit. For example, NGC\,2602 (age~$\approx 35$~Myr) and NGC\,1039 (age~$\approx 132$~Myr) have no MSTO stars with $v \sin i<30$. In contrast, Collinder\,463 (age~$\approx 281$~Myr) has $f_{\rm slow\,rot}^{v \sin i<30} = $~0.21 and has a bimodal $v \sin i$ distribution. In addition to tidal synchronization in binaries as reported by \citet{Sun2019b} for a similar age cluster NGC\,2287, star-disk interaction during the pre-main-sequence stage may also have contributed to the slow-rotator population in this cluster. 

The estimated median value of $f_{\rm slow,rot}^{v \sin i < 100}$ in this work is consistent with previous determinations of slow-rotator fractions in LMC clusters. For example, \citet{Kamann2020} reported a fast-rotator fraction of $0.37$--$0.58$ across different mass ranges in the $\sim1.5$~Gyr LMC cluster NGC\,1846, implying a corresponding slow-rotator fraction of $f_{\rm slow,rot} \approx 0.42$--$0.63$. By comparing observed CMD with Monte Carlo simulations of synthetic cluster \citet{Correnti2021MNRAS.504..155C} estimated that $\sim40\%$ of stars in the LMC cluster NGC\,1831 are slow rotators. Additionally, using $v \sin i$ measurements given by \citet{Kamann2025MNRAS.542.2768K}, we derived $f_{\rm slow,rot}^{v \sin i < 100} $ as $\approx 0.39$ and $\approx 0.28$ for young (200--300~Myr) LMC clusters NGC\,1866 and NGC\,1856, respectively. Additionally, \citet{Leanza2025A&A...698A..27L} reported that $\approx 18 \%$ of MSTO stars of NGC\,1783 LMC cluster (age~$=1.5$~Gyr) have $v \sin i < 50$~km~s$^{-1}$.

Additionally, $f_{\rm slow\, rot}$ is plotted against the binary fractions and $log{\rm (age/yr)}$ as illustrated in Fig~\ref{fig:slow_rot}. Here, Class I, II, and III OCs are divided into two age groups to mitigate the effect of the correlation between age and the binary fraction, as shown in Fig.~\ref{fig:fb_logage}. Due to the lack of $v \sin i$ measurements for all MSTO stars, OCs that have at least 50\% MSTO stars with $v \sin i$ measurements are overplotted. None of the observed correlations are statistically significant, as assessed using Spearman's rank correlation coefficient and Pearson correlation coefficient. However, a mild increasing correlation is observed between the fraction of slow rotators and the binary fraction for OCs with $\log g{\rm (age/yr)} > 8$. This may indicate the effect of tidal synchronization in binaries, because it is a cumulative process whose efficiency increases with time. However, the tidal synchronization in binaries is dependent on their orbital and dynamical properties as well as the cluster environment. The statistically robust detection of this trend requires larger sample clusters with complete $v \sin i$ measurements and total binary fractions.

\subsubsection{Role of binaries}

Tidal synchronization and subsequent evolution to slow rotators in binaries depend on their orbital parameters and mass ratios. Binaries with $P \leq 10$--20~days are expected to be tidally circularized \cite{Zahn1977A&A....57..383Z}.  Recently, \citet{Li2020} using 45 Doradus stars in eclipsing binaries demonstrated that the majority of these binaries with periods below 10~days have core rotations synchronized with their orbital periods, resulting in slow rotation. Some systems even show rotation rates slower than synchronous, suggesting that unstable, tidally excited oscillations can transfer angular momentum from the star to the orbit, leading to sub-synchronous rotation.

Our sample includes known eclipsing or spectroscopic binaries, which allow us to investigate the impact of stellar companions on stellar rotation.  For this, we exploited the variability catalogs provided by \textit{Gaia} DR3 \citep{Eyer2023A&A...674A..13E} and \citet{Gavras2023A&A...674A..22G} of MSTO members. These binaries are shown as black and cyan open circles in Figs~\ref{fig:split_ms}, \ref{fig:eMSTO}, \ref{fig:weird_ocs}, and \ref{fig:old_ocs}, and their parameters are listed in Table~\ref{tab:binaries}. 
The correlations between $log(P/{\rm days})$, $v \sin i$ (when available), and eccentricities (when available) are shown in Fig~\ref{fig:bin_var}. These binaries have $P = 0.30$--3636~days, $v \sin i = 7$--243~km~s$^{-1}$, and $e = 0$--0.54. Although their CMD positions vary in different clusters, most of these binaries occupy the redder regions of the CMDs.

Of 30 binaries with available $v \sin i$ measurements,  11 have $v \sin i \leq 30$~km~s$^{-1}$, and all but two are slow to moderate rotators with $v \sin i \leq 150 $~km~s$^{-1}$. No correlation between $log(P/{\rm days})$ and $v \sin i$ is observed. An increasing correlation between $v \sin i$ and $e$ is observed except for two binaries with large $v \sin i$ but $e = 0$. One of the two binaries (Gaia~DR3~5614817313079742464) is a spectroscopic binary and has a period of 113.797~days. Its large $v \sin i$ and circular orbit indicate that the binary has circularized but not yet tidally synchronized, which is expected given its long period. 
The other binary (Gaia~DR3~5290847380178313856) also a spectroscopic binary, with $P= 5.099$~days and $e=0$, is a magnetically active variable \citep[ACV/CP/MCP/ROAM/ROAP/SXARI-type variable;][]{Eyer2023A&A...674A..13E}, with a rotation period of 0.4296~days \citep{Bouma2021}. Based on the fitted isochrone, its primary has a mass of 2.5~M$_{\odot}$, and is located on the bluer edge of the CMD (see Fig~\ref{fig:eMSTO}). These properties indicate that the binary may have undergone stellar interactions or accretion, making it bluer, brighter, and a fast rotator. Such characteristics are typical of blue stragglers and blue lurkers formed via accretion from companions \citep{Ferraro2023}.

In the $log P$-$e$ plane, a few binaries with $e > 0.1$ with $P < 20$~d are observed. These binaries have high $v \sin i$, indicating that they have not yet reached the tidal circularization stage. Similar cases have been reported for highly eccentric late B-type binaries, consistent with very young ages and not yet having reached the tidal synchronization stage \citep{Rucinski2007MNRAS.380L..63R,Maceroni2009A&A...508.1375M}. 

Several OCs, including NGC\,2287 \citep{Sun2019b}, NGC\,2355 \citep{Maurya2024MNRAS.532.1212M}, NGC\,2422 \citep{He2023MNRAS.525.5880H}, NGC\,2423 \citep{Bu2024ApJ...968...22B} NGC\,3532 \citep{He2025ApJ...979..246H,Rao2025arXiv251205458R}, NGC\,6067 \citep{Maurya2025ApJ...989..123M} also reported that the slow-rotator population cannot be fully explained by tidal synchronization in binaries and requires additional mechanisms.

\subsubsection{Role of variables}

\begin{figure}
    \centering
    \includegraphics[width=1\linewidth]{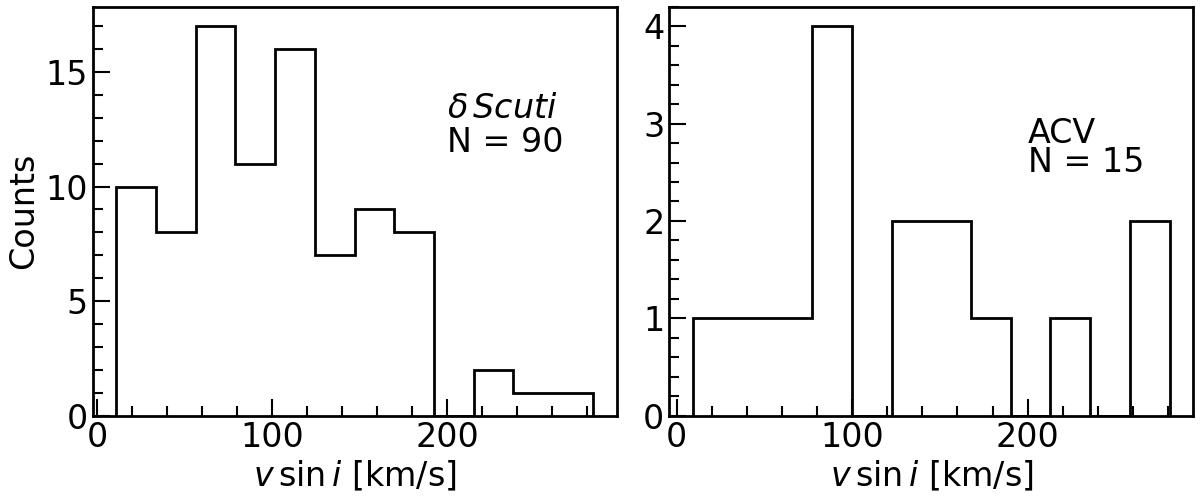}
   \caption{$v \sin i$ distribution of DSCT$|$GDOR$|$SXPH and ACV$|$CP$|$MCP$|$ROAM$|$ROAP$|$SXARI types variables.}
    \label{fig:var}
\end{figure}

We now examine the $v \sin i$ distribution of pulsating and chemically peculiar variables.  In total, we identified 645 sources classified as DSCT ($\delta$ Scuti), GDOR ($\gamma$ Doradus), or SXPH (SX Phoenicis) and 47 as ACV ($\alpha^2$~Canum~Venaticorum), CP (Chemically Peculiar), MCP (Magnetic Chemically Peculiar), ROAM (Rapidly Oscillating Am), ROAP (Rapidly Oscillating Ap), or SXARI (SX Arietis) type variables in our OC sample \citep{Eyer2023A&A...674A..13E}. Among these, 90 DSCT$|$GDOR$|$SXPH and 15 ACV$|$CP$|$MCP$|$ROAM$|$ROAP$|$SXARI have $v \sin i$ measurements available. The resulting $v \sin i$ distributions of these variables are shown in Fig.~\ref{fig:var}. 

For pulsating variables, the mass range according to fitted isochrones is 1.0--2.5~M$_{\odot}$, which corresponds to a maximum of $v \sin i \approx 100$--300~km~s$^{-1}$.  However, the observed distribution is dominated by slow-to-moderate rotators, with only a small fraction of MSTO stars with rapid rotation. DSCT and GDOR variables generally have high $v \sin i$  \citep{Gootkin2024,Wang2025}, but they often remain undetected 
because of less coherent periodic behavior \citep{Murphy2024}. Consequently, the detection biases affecting DSCT and GDOR variables make it difficult to assess their contribution to the slow-rotator population.

Chemically peculiar stars (Ap/Bp spectral type) typically exhibit strong, stable magnetic fields ($\sim200$~G to $\sim3$~kG) and anomalous surface abundances \citep{Braithwaite2004Natur.431..819B, Keszthelyi2022MNRAS.517.2028K}. These types of variables are generally slow to moderate rotators \citep{Netopil2017, Bauer2024}. From Fig~\ref{fig:var}, we observed that these variables display a nearly flat $v \sin i$ distribution, spanning 9--281~km~s$^{-1}$.The broader range seen in our sample may reflect a mixture of different subtypes and inclination effects. We note that our chemically peculiar stars sample is likely incomplete, since the identification of these stars requires high-resolution spectroscopic abundance analysis, and the \textit{Gaia} variability catalog preferentially detects photometrically prominent cases only. Therefore, given the relatively small number of these stars in our sample, we caution against drawing firm conclusions.  

\section{Summary and Conclusion} \label{sec:summary}

We conducted a systematic investigation of eMS and eMSTO phenomena in 53 Galactic OCs with ages of 10~Myr to 2.4~Gyr using $v \sin i$ measurements from \textit{Gaia} DR3 and GES data. Based on the extinction, MSTO morphologies, and distribution of fast and slow rotators, we divided clusters into 4 classes. We find a strong dependence of the detectability of split MS features on cluster extinction. Namely, OCs with little extinction ($A_{\rm v} < 0.15$; Class~I) exhibit clear photometric separation between fast and slow rotators, manifesting as well-defined split MS. These OCs are defined as a golden sample and are most suitable for direct comparison with rotating stellar evolutionary models \citep{Nguyen2022, Nguyen2025A&A...701A.258N} and for investigating the intrinsic formation mechanisms of fast and slow rotators. Class~II OCs ($ 0.15 < A_{\rm v} < 0.5$), on the other hand, display apparent eMSTO features but with moderate to no distinction in fast and slow rotators on MSTO. Even after differential reddening correction, residual extinction effects may still displace stars in the CMD; therefore, caution must be exercised when interpreting the origin of fast and slow rotators in these clusters based solely on photometric and spatial information. Class~III OCs show no apparent eMS or eMSTO features due to very sparse upper MS populations, primarily as a consequence of significant dynamical mass loss rather than an absence of intrinsic rotational bimodality. Class~IV OCs probe the empirical upper age limit of the eMSTO phenomenon and the onset of magnetic braking among intermediate-mass stars. NGC\,2420 ($\log{\rm(age/yr)} = 9.335$) and NGC\,2141 ($\log{\rm(age/yr)}= 9.33$), which are sub-solar ($[Fe/H] \leq -0.05$~dex) with turnoff masses of $\sim 1.5~M_\odot$, are dominated by slow rotators, consistent with efficient magnetic braking having already operated at this mass and age. On the other hand, Trumpler\,20 
($\log{\rm(age/yr)} = 9.3$ and $[Fe/H] = 0.15$~dex) have still not experienced a 
significant magnetic braking.

The MSTO width increases with cluster age, albeit with considerable scatter, consistent with the observational results of \citet{Niederhofer2015MNRAS.453.2070N}, \citet{Cordoni2018}, and \citet{Cordoni2024}, and with the theoretical predictions of rotating stellar models \citep{Georgy2019A&A...622A..66G, Brandt2015ApJ...807...25B}. The dispersion in this correlation is mainly driven by differences in OCs' extinction, where higher extinction OCs exhibit systematically larger MSTO color spreads at any given age. This indicates that the effect of extinction must be accounted for when comparing observations with predictions from rotating stellar evolutionary models. We find no significant dependence of MSTO width on cluster mass. However, mass estimates are currently available for only a limited subset of our sample, which restricts the statistical significance of this result. A larger and more complete dataset will be essential to robustly evaluate the role of cluster mass in shaping MSTO morphology. 

We estimated $f_{\rm slow \, rot}$ for the largest homogeneous sample of Galactic OCs to date, spanning a wide range of ages, extinctions, and Galactic environments. $f_{\rm slow\, rot}^{v \sin i<100}$ and $f_{\rm slow\, rot}^{v \sin i<30}$ remain roughly constant with a mild increase for older clusters. The estimated median value of $f_{\rm slow\, rot}^{v \sin i<100} \approx 0.41$, suggesting that a substantial fraction of intermediate-mass stars either arrive on the main sequence already rotating slowly/moderately or undergo spin-down on timescales shorter than the youngest clusters in our sample. This result is consistent with intrinsic bimodality reported by \citet{Cordoni2024} for several galactic clusters, and LMC clusters \citep{Correnti2021MNRAS.504..155C, Kamann2020, Kamann2023MNRAS.518.1505K, Kamann2025MNRAS.542.2768K, Leanza2025A&A...698A..27L}. The fraction of stars reaching near-zero rotation $f_{\rm slow\, rot}^{v \sin i<30} \approx 0.08 $ is much lower, indicating that a majority of intermediate-mass stars may never reach the complete spin-down limit on MS. A slight positive correlation with binary fraction suggests that binaries may contribute to slow rotators through tidal synchronization, however it is dependent on several other factors like cluster environment, orbital and dynamical parameters of binaries. Using $v \sin i$, $P$, and $e$ distribution of 52 known MSTO binaries, with 30 available with $v \sin i$ measurements, we find that the majority of the close binaries are slow to moderate rotators. The increasing correlation between $v \sin i$ and eccentricity is consistent with the expectation that tidally circularized systems rotate more slowly. Though binaries may lead to slow rotation, the very close binaries can even start interacting and become fast rotators, as we observed for one binary in our sample. 

The pulsating variable population is dominated by slow-to-moderate rotators, though detection biases arising from less coherent periodic behavior \citep{Murphy2024} make it difficult to quantify their intrinsic contribution to the slow-rotator population. Chemically peculiar stars (Ap/Bp type) are typically slow rotators due to their strong magnetic fields \citep{Braithwaite2004Natur.431..819B, Keszthelyi2022MNRAS.517.2028K}, but their observed population of ACV type variable in our sample is far too low, likely due to incompleteness, to account for the measured slow-rotator fractions. 

Our classification framework, and in particular the identification of the Class~I golden sample, provides a physically motivated basis for future studies to isolate and test different spin-down mechanisms and calibrate stellar rotation models. Future large-scale time series and spectroscopic data, such as \textit{Gaia} DR4 and LSST, will provide variability and rotational velocity information for more intermediate-mass stars across a wide range of cluster ages and Galactic environments. This will allow us to disentangle contributions of different spin-down mechanisms and provide the observational constraints needed to calibrate the next generation of rotating stellar evolutionary models.

\begin{acknowledgments}
    We thank the anonymous referee for constructive feedback. 
    KKR and WPC acknowledge funding from the National Science and Technology Council of Taiwan (NSTC~113-2123-M-008-004). This work used the third data release from the European Space Agency (ESA) mission {\it Gaia} (\url{https://www.cosmos.esa.int/gaia}), \citep{GaiaDR32023}, processed by the \textit{Gaia} Data Processing and Analysis Consortium (DPAC, \url{https://www.cosmos.esa.int/web/gaia/dpac/consortium}). This research utilised the Astrophysics Data System (ADS), governed by NASA (\url{https://ui.adsabs.harvard.edu}).
    The following tools are used for analyses carried out in this work: 
    \textsc{Astropy} \citep{2013A&A...558A..33A}; 
    \textsc{Astroquery} \citep{Ginsburg2019AJ....157...98G};
    \textsc{Matplotlib} \citep{Hunter:2007};
    \textsc{NumPy} \citep{2020Natur.585..357Harris}; 
    \textsc{SciPy} \citep{2020SciPy-NMeth}; 
    \textsc{topcat} \citep{2005ASPC..347...29TOPCAT}. 

\end{acknowledgments}

\appendix

\begin{table}[!ht]
    \centering
        \caption{The binary fractions and slow rotator fractions of the 53 OCs. Here,  Column 1: cluster name;  Columns 2, 3, and 3: limiting isochrone, transition isochrone, and transition magnitude used to estimate binary fractions;  Column 4: binary fraction;  Columns 5 and 6: number of MSTO stars and MSTO spreads;  Columns 7, 8, and 9: fraction of MSTO stars with available $v \sin i $ measurements, with $0< v \sin i < 100$~km~s$^{-1}$, and with $0< v \sin i < 30$~km~s$^{-1}$ .}
	\label{tab:bin_frac}
 {\fontsize{8pt}{8.2pt} \selectfont
    \begin{tabular}{cccccccccc}
            \hline
            \\
        cluster & No of MSTO stars & MSTO$_{\rm spread}$ & iso$_{q_{\rm lim}}$ & iso$_{q_{\rm trans}}$ & Gmag$_{\rm trans}$ & $f_b^{q>q_{\rm lim}}$  &$f_{\rm MSTO stars}^{v \sin i>0}$ & $f_{\rm MSTO stars}^{0<v \sin i<100}$ & $f_{\rm MSTO stars}^{0<v \sin i<30}$ \\
        \\
        \hline
        Alessi 1 & 9 & 0.075$\pm$ 0.03 & 0.9 & 0.9 & 17.5 & 0.26 & 1 & 0.8 & 0.1 \\ 
        Alessi 6 & 16 & 0.184$\pm$ 0.053 & 0.8 & 0.8 & 17.5 & 0.26 & 0.55 & 0.43 & 0.14 \\ 
        ASCC 113 & 18 & 0.08$\pm$ 0.013 & 0.7 & 0.9 & 17.5 & 0.14 & 0.86 & 0.38 & 0.03 \\ 
        Collinder 463 & 20 & 0.224$\pm$ 0.033 & 0.85 & 0.8 & 16 & 0.26 & 0.57 & 0.47 & 0.21 \\ 
        IC 2602 & 21 & 0.07$\pm$ 0.017 & 0.82 & 0.7 & 16 & 0.13 & 0.57 & 0.25 & 0 \\ 
        IC 4665 & 10 & 0.121$\pm$ 0.067 & 0.8 & 1 & 17 & 0.21 & 0.67 & 0.58 & 0.08 \\ 
        Melotte 22 & 14 & 0.086$\pm$ 0.012 & 0.7 & 0.6 & 17 & 0.15 & 0.74 & 0.26 & 0.09 \\ 
        NGC 1039 & 15 & 0.06$\pm$ 0.06 & 0.65 & 0.8 & 17.6 & 0.2 & 0.55 & 0.5 & 0 \\ 
        NGC 1901 & 12 & 0.028$\pm$ 0.036 & 0.85 & 0.9 & 17 & 0.26 & 0.4 & 0.75 & 0.25 \\ 
        NGC 1912 & 38 & 0.106$\pm$ 0.014 & 0.8 & 0.8 & 17.5 & 0.2 & 0.14 & 0.32 & 0.08 \\ 
        NGC 2099 & 174 & 0.139$\pm$ 0.011 & 0.8 & 0.8 & 17.5 & 0.2 & 0.09 & 0.41 & 0.21 \\ 
        NGC 2141 & 129 & 0.076$\pm$ 0.006 & 0.9 & 0.9 & 18.5 & 0.24 & 0.63 & 0.99 & 0.39 \\ 
        NGC 2158 & 153 & 0.125$\pm$ 0.009 & 0.9 & 0.9 & 18 & 0.25 & 0.2 & 0.81 & 0.24 \\ 
        NGC 2168 & 25 & 0.131$\pm$ 0.028 & 0.8 & 0.8 & 17.5 & 0.22 & 0.14 & 0.47 & 0.19 \\ 
        NGC 2287 & 25 & 0.06$\pm$ 0.012 & 0.6 & 0.9 & 17.3 & 0.16 & 0.96 & 0.47 & 0.05 \\ 
        NGC 2301 & 6 & 0.101$\pm$ 0.023 & 0.7 & 0.9 & 17.6 & 0.15 & 0.34 & 0.5 & 0.24 \\ 
        NGC 2360 & 86 & 0.11$\pm$ 0.019 & 0.8 & 0.9 & 18.6 & 0.18 & 0.48 & 0.41 & 0.1 \\ 
        NGC 2420 & 77 & 0.029$\pm$ 0.003 & 0.86 & 0.9 & 18.5 & 0.14 & 0.71 & 1 & 0.62 \\ 
        NGC 2422 & 7 & 0.075$\pm$ 0.018 & 0.8 & 1 & 17.5 & 0.13 & 0.73 & 0.45 & 0.12 \\ 
        NGC 2423 & 56 & 0.104$\pm$ 0.016 & 0.8 & 0.6 & 16 & 0.18 & 0.69 & 0.5 & 0.17 \\ 
        NGC 2437 & 97 & 0.113$\pm$ 0.008 & 0.8 & 0.8 & 17 & 0.16 & 0.08 & 0.5 & 0.24 \\ 
        NGC 2447 & 54 & 0.082$\pm$ 0.015 & 0.7 & 0.7 & 16 & 0.18 & 0.68 & 0.4 & 0.08 \\ 
        NGC 2451B & 2 & 0.09$\pm$ 0.045 & 0.9 & 0.8 & 15 & 0.14 & 0.59 & 0.15 & 0.08 \\ 
        NGC 2482 & 18 & 0.085$\pm$ 0.017 & 0.7 & 0.7 & 15 & 0.18 & 0.32 & 0.08 & 0 \\ 
        NGC 2516 & 68 & 0.095$\pm$ 0.013 & 0.7 & 0.7 & 17.5 & 0.17 & 0.73 & 0.43 & 0.07 \\ 
        NGC 2527 & 26 & 0.125$\pm$ 0.04 & 0.8 & 1 & 18 & 0.15 & 0.86 & 0.42 & 0.08 \\ 
        NGC 2539 & 46 & 0.103$\pm$ 0.021 & 0.7 & 0.7 & 18 & 0.17 & 0.36 & 0.39 & 0.13 \\ 
        NGC 2548 & 32 & 0.09$\pm$ 0.022 & 0.7 & 0.9 & 18 & 0.15 & 0.84 & 0.47 & 0.14 \\ 
        NGC 2632 & 16 & 0.113$\pm$ 0.04 & 0.7 & 1 & 15.7 & 0.19 & 0.68 & 0.4 & 0.12 \\ 
        NGC 3114 & 49 & 0.064$\pm$ 0.012 & 0.7 & 0.7 & 15.7 & 0.17 & 0.13 & 0.39 & 0.16 \\ 
        NGC 3293 & 73 & 0.06$\pm$ 0.008 & 1 & 0.9 & 14 & 0.17 & 0.59 & 0.24 & 0.01 \\ 
        NGC 3532 & 55 & 0.107$\pm$ 0.013 & 0.6 & 1 & 17.2 & 0.22 & 0.47 & 0.38 & 0.12 \\ 
        NGC 3766 & 83 & 0.06$\pm$ 0.012 & 0.75 & 0.95 & 16 & 0.17 & 0.63 & 0.19 & 0.04 \\ 
        NGC 5460 & 10 & 0.09$\pm$ 0.02 & 0.7 & 0.7 & 15.7 & 0.23 & 0.71 & 0.33 & 0.08 \\ 
        NGC 5822 & 75 & 0.163$\pm$ 0.021 & 0.7 & 0.7 & 15.7 & 0.19 & 0.4 & 0.41 & 0.07 \\ 
        NGC 6005 & 98 & 0.111$\pm$ 0.012 & 0.9 & 0.9 & 18.5 & 0.2 & 0.21 & 0.39 & 0.11 \\ 
        NGC 6025 & 14 & 0.071$\pm$ 0.02 & 0.8 & 0.6 & 15.8 & 0.2 & 0.43 & 0.7 & 0.3 \\ 
        NGC 6067 & 79 & 0.085$\pm$ 0.013 & 0.8 & 0.8 & 19 & 0.18 & 0.08 & 0.24 & 0.07 \\ 
        NGC 6281 & 10 & 0.122$\pm$ 0.023 & 0.7 & 0.8 & 17.4 & 0.22 & 0.54 & 0.38 & 0.07 \\ 
        NGC 6475 & 17 & 0.105$\pm$ 0.023 & 0.7 & 1 & 16 & 0.16 & 0.5 & 0.52 & 0.15 \\ 
        NGC 6633 & 30 & 0.132$\pm$ 0.034 & 0.8 & 1 & 17.2 & 0.28 & 0.66 & 0.41 & 0.17 \\ 
        NGC 6649 & 118 & 0.175$\pm$ 0.018 & 0.9 & 0.6 & 16.7 & 0.09 & 0.13 & 0.44 & 0.12 \\ 
        NGC 6705 & 143 & 0.102$\pm$ 0.009 & 0.8 & 0.8 & 18.5 & 0.21 & 0.22 & 0.3 & 0.06 \\ 
        NGC 6802 & 65 & 0.14$\pm$ 0.02 & 0.8 & 0.8 & 18.5 & 0.23 & 0.17 & 0.35 & 0.1 \\ 
        NGC 6811 & 32 & 0.132$\pm$ 0.02 & 0.7 & 0.7 & 17.2 & 0.24 & 0.43 & 0.52 & 0.14 \\ 
        NGC 6940 & 64 & 0.154$\pm$ 0.017 & 0.8 & 0.8 & 16 & 0.21 & 0.55 & 0.57 & 0.18 \\ 
        NGC 7209 & 41 & 0.222$\pm$ 0.037 & 0.8 & 0.7 & 17.2 & 0.22 & 0.49 & 0.28 & 0.11 \\ 
        Pismis 15 & 32 & 0.151$\pm$ 0.029 & 0.9 & 0.9 & 18.5 & 0.16 & 0.35 & 0.54 & 0.08 \\ 
        Roslund 6 & 7 & 0.082$\pm$ 0.027 & 0.6 & 0.7 & 18 & 0.11 & 0.85 & 0.41 & 0.12 \\ 
        Ruprecht 134 & 78 & 0.161$\pm$ 0.018 & 1 & 1.2 & 18.5 & 0.25 & 0.41 & 0.28 & 0.05 \\ 
        Trumpler 10 & 18 & 0.058$\pm$ 0.034 & 0.8 & 1 & 17.3 & 0.16 & 0.74 & 0.34 & 0.03 \\ 
        Trumpler 20 & 235 & 0.091$\pm$ 0.007 & 0.9 & 1.1 & 18.5 & 0.05 & 0.51 & 0.75 & 0.34 \\ 
        Trumpler 23 & 74 & 0.164$\pm$ 0.021 & 1 & 1 & 18.5 & 0.22 & 0.18 & 0.43 & 0.03 \\
        \\
        \hline
    \end{tabular}}
\end{table}

\begin{table}[!ht]
    \centering
        \caption{Parameters of the MSTO binaries.}
	\label{tab:binaries}
 {\fontsize{6pt}{8pt} \selectfont

    \begin{tabular}{cccccccccccc}
        \hline
        \\
        cluster & \textit{Gaia} DR3 ID & RA & DEC & G & BP$-$RP & $v \sin i$ & RUWE & Type & Period & $e$ & REF$_{v \sin i}$ \\
        
         &  & (deg) & (deg) & (mag) & (mag) & (km~s$^{-1}$) &  & & (days) &  &  \\ 
        \\
        \hline
        \\
        Melotte 22 & 65207709611941376 & 56.851821 & 23.914478 & 7.286 & 0.042 & 7$\pm$3 & 1.001 & SB & 7.3452 & 0 & 1 \\ 
        Melotte 22 & 66507469798631808 & 57.491988 & 23.848487 & 6.806 & 0.055 & 9$\pm$2 & 0.974 & SB & 16.726 & 0 & 1 \\ 
        NGC 3532 & 5337851944644789632 & 167.090170 &  $-$59.556335 & 9.597 & 0.121 & 17.9$\pm$8.0 & 0.879 & SB & 6.928 & 0 & 2 \\ 
        IC 4665 & 4474066504530306688 & 266.3902198 & 5.715654 & 7.659 & 0.055 & 25 & 1.024 & SB & 10.526 & 0$\pm$0.01 & 3 \\ 
        \\
        \hline
    \end{tabular}}
{\scriptsize $^1$\citet{Torres2021ApJ...921..117T}, $^2$\citet{GaiaDR32023}, $^3$\citet{Glcebocki2005}}

Only the first 4 rows of the table are shown here. The complete table will be available in machine-readable form in the online version of the paper.
\end{table}


\bibliography{references}{}
\bibliographystyle{aasjournal}

\end{document}